\documentclass[useAMS,usenatbib]{mn2e}
\usepackage{graphicx}
\usepackage{epsfig}
\usepackage{multirow}
\usepackage{float}

\title[XMM-Newton Observation of An Extreme NLS1: PG 1244+026]{A Long XMM-Newton Observation of An Extreme Narrow Line Seyfert 1: PG 1244+026}

\author[C. Jin, C. Done M. Middleton,  M. Ward]
{Chichuan Jin$^{1}$, Chris Done$^1$, Matthew Middleton$^{1,2}$, Martin Ward$^{1}$\\
$^{1}$Department of Physics, University of Durham, South Road, Durham, DH1 3LE, UK\\
$^{2}$Astronomical Institute, Anton Pannekoek, University of Amsterdam, Science Park 904, 1098XH, Amsterdam, The Netherlands}

\begin{document}

\maketitle

\label{firstpage}

\begin{abstract}

We explore the origin of the strong soft X-ray excess in Narrow Line
Seyfert 1 galaxies using spectral-timing information from a 120ks {\it
XMM-Newton} observation of PG 1244+026. Spectral fitting alone cannot
distinguish between a true additional soft X-ray continuum component
and strongly relativistically smeared reflection, but both models also
require a separate soft blackbody component. This is most likely
intrinsic emission from the disc extending into the lowest energy
X-ray bandpass.
The {\it RMS} spectra on short timescales (200-5000s) contain
both (non-disk) soft excess and power law emission. However, the
spectrum of the variability on these timescales correlated with the
4-10~keV lightcurve contains only the power law. Together these show
that there is fast variability of the soft excess which is independent
of the 4-10~keV variability. This is inconsistent with a single
reflection component making the soft X-ray excess as this necessarily
produces correlated variability in the 4-10~keV bandpass. Instead, the
{\it RMS} and covariance spectra are consistent with an additional
cool Comptonisation component which does not contribute to the
spectrum above 2~keV.

\end{abstract}


\section{Introduction}

The spectral energy distribution (SED) of Active Galactic Nuclei (AGN)
is powered by a mass accretion rate, $\dot{M}$, onto a central
super-massive black hole of mass $M\sim 10^{6-9}M_\odot$. To the first-order
approximation, as seen in the stellar mass black hole binaries, the state of
the accretion flow is determined by $L/L_{Edd}$, where $L=\eta \dot{M}
c^2$, $\eta$ is the efficiency set by black hole spin and the nature
of the accretion flow, and
$L_{Edd}=1.3\times 10^{38}M$ is the Eddington limit.

The highest Eddington ratio accretion flows, with $L/L_{Edd}\sim 1$,
are found in the subset of broad line AGN known as Narrow-Line Seyfert
1s (NLS1s, e.g.  \citealt{Leighly99}) which have permitted line widths
only slightly broader than those of their forbidden lines
(\citealt{Osterbrock85}, Boroson \& Green 1992), indicating relatively
low black hole masses ($10^{6-7}M_\odot$: Boroson 2002). This
combination of properties means that their accretion disc spectra
should peak in the EUV/soft X-ray bandpass rather than the far UV peak
expected for more typical Broad Line Seyfert 1 (BLS1)/QSOs
which have $10^8M_\odot$, $L/L_{Edd}\sim 0.1$.

However, it has long been known that the spectra of AGN are more
complex than expected from simple disc models. Standard QSO template
spectra show substantial hard and soft X-ray emission as well as a
`big blue bump' from a standard disc (Elvis et al. 1994; Richards et al
2006). There is a power law which dominates in the 2-10~keV bandpass,
and a soft X-ray excess over the low energy extrapolation of the power
law emission which appears to be ubiquitous in all AGN. The power law
`coronal' emission is commonly seen also in black hole binaries (BHB),
but the soft X-ray excess has no clear counterpart in the stellar mass
systems. This could be a true additional continuum component seen only
in AGN (Laor et al. 1997; Magdziarz et al. 1998; Gierli\'{n}ski \& Done
2004), but the characteristic temperature of this component remains
remarkably stable across objects of very different mass (Czerny et al.
2003; Gierli\'{n}ski \& Done 2004), making this solution fine-tuned. The
only current alternative model in the literature is that the soft
excess instead arises as result of reflection and reprocessing from
partially ionised material (Crummy et al. 2006; Walton et al. 2013). The
decrease in opacity in this material below the Oxygen K and iron L
edges at $\sim 0.7$~keV give a physical reason for the fixed energy of
this feature. However, this requires similar fine tuning of the
ionisation state of the reflecting material in order to always be
dominated by the opacity at $\sim 0.7$~keV (Done \& Nayakshin 2007).
Also, the strong soft X-ray lines predicted by this model are not seen
in the data, requiring extreme relativistic effects (high spin and
highly centrally concentrated illumination) to smear these into a
pseudo-continuum (Crummy et al. 2006; Walton et al. 2013).

These two very different ideas for the origin of the soft X-ray excess
can be tested via variability. In the simplest reflection
interpretation, both soft and hard X-rays are connected together by a
single, partially ionised reflection component. Instead, if the soft
X-ray excess is a true additional continuum component, this does not
extend into the hard X-ray band and the soft and hard X-rays are
decoupled.  Previous work has addressed this via a range of different
techniques mostly aimed at separating the spectrum into constant and
variable components.  In particular, the variable component can be
quantified by calculating the excess variance in each (binned) energy
channel (Edelson et al. 2002). These {\it RMS} spectra reveal a
bewildering range of shapes (e.g. the compilation of Gierli\'{n}ski \&
Done 2006). These appear to be correlated with complexity of the X-ray
spectra. The {\it RMS} spectra become more uniform after excluding
NLS1 which show deep X-ray minima (these are the spectra which can be
interpreted as being dominated by extreme relativistic reflection:
Gallo 2006), and all objects where there is a strong warm
absorber. Then, the remaining `simple' AGN show {\it RMS} spectra
where the low energy variability is strongly suppressed. These are
consistent with a two component spectrum, where there is a stable soft
component dominating at low energies, and a variable power law which
dominates at high energies. This is seen in both NLS1 (Middleton et al.
2009; Jin et al. 2009) and BLS1 (Mehdipour et al. 2011; Noda et al. 2011;
2013).

Here we use an archetypal `simple' NLS1 PG 1244+026 to further test these ideas
on the origin of the soft X-ray excess and the shape of the SED. PG 1244+026 was
selected by Jin et al. (2012a) (hereafter: J12a) as one of the brightest
un-obscured Type 1 AGN with both {\it XMM-Newton} and {\it SDSS} spectra.  From
an analysis of optical spectra, PG 1244+026 can also be seen to have the
narrowest H$\beta$ line (830 km s$^{-1}$) and the fourth smallest
H$\beta$ 
equivalent width (41\AA) of all of the PG quasars (Boroson \& Green
1992). Previous X-ray observations of this source with {\it ASCA} revealed
possible complexity around oxygen/iron L shell energies (\citealt{Fiore98};
\citealt{Ballantyne01}), but the data were limited. Here we report on a new
120ks {\it XMM-Newton} observation which allows us to study the spectrum and
variability of the source in detail.

This paper is organized as follows. We first describe the data quality and our
data reduction procedure for this new {\it XMM-Newton} observation. Then in
section 3 we try various models to fit the 0.3-10 keV spectra and report the
results. Section 4 will focus on the variability of PG 1244+026, especially the
light-curve, power spectral density (PSD) and the frequency-differentiated {\it
  RMS} spectra. In section 5, we will present the frequency-dependent covariance
spectra and discuss their properties. In section 6, we will discuss issues such
as broadband SED and black hole mass.
The summary and conclusions are in Section 7.

%%%%%%%%%%%%%%%%%%%%%%%%%%%%%%%%%%%%%%%%%%%%%%%%%%%%%%%%%%%%%%%%%%%%%%%%%%%%%%%%%%%%%
\begin{table*}
 \centering
  \begin{minipage}{175mm}
   \caption{The model expression in {\sc xspec} v12.7.1 for the four spectral
     fitting methods in Figure~\ref{spec-fit-fig}.}
     \begin{tabular}{@{}lll@{}}
\hline
Model Name & Model Expression in {\sc xspec} v12.7.1 & seed photons \\
\hline
{\tt COMP-BBODY} & {\tt CONSTANT*WABS*ZWABS*( BBODY + NTHCOMP +
  COMPTT + KDBLUR*PEXMON )} & $kT_{bb}$ in {\sc bbody} for {\sc
  comptt} and {\sc nthcomp}\\
{\tt COMP-COMPTT} & {\tt CONSTANT*WABS*ZWABS*( BBODY + NTHCOMP +
  COMPTT + KDBLUR*PEXMON )} & $kT_e$ in {\sc comptt} for {\sc
  nthcomp} \\
{\tt REFL} & {\tt CONSTANT*WABS*ZWABS*( BBODY + NTHCOMP +
  KDBLUR*RFXCONV*NTHCOMP )} & $kT_{bb}$ in {\sc bbody} for {\sc nthcomp}\\
{\tt IONPCF} & {\tt CONSTANT*WABS*ZWABS*ZXIPCF*( BBODY +
  NTHCOMP )} & $kT_{bb}$ in {\sc bbody} for {\sc nthcomp}\\
\hline
     \end{tabular}
 \end{minipage}
 \label{spec-model-tab}
\end{table*}
%%%%%%%%%%%%%%%%%%%%%%%%%%%%%%%%%%%%%%%%%%%%%%%%%%%%%%%%%%%%%%%%%%%%%%%%%%%%%%%%%%%%%

\begin{figure*}
\begin{center}
\begin{tabular}{l}
 \epsfxsize=14cm \epsfbox{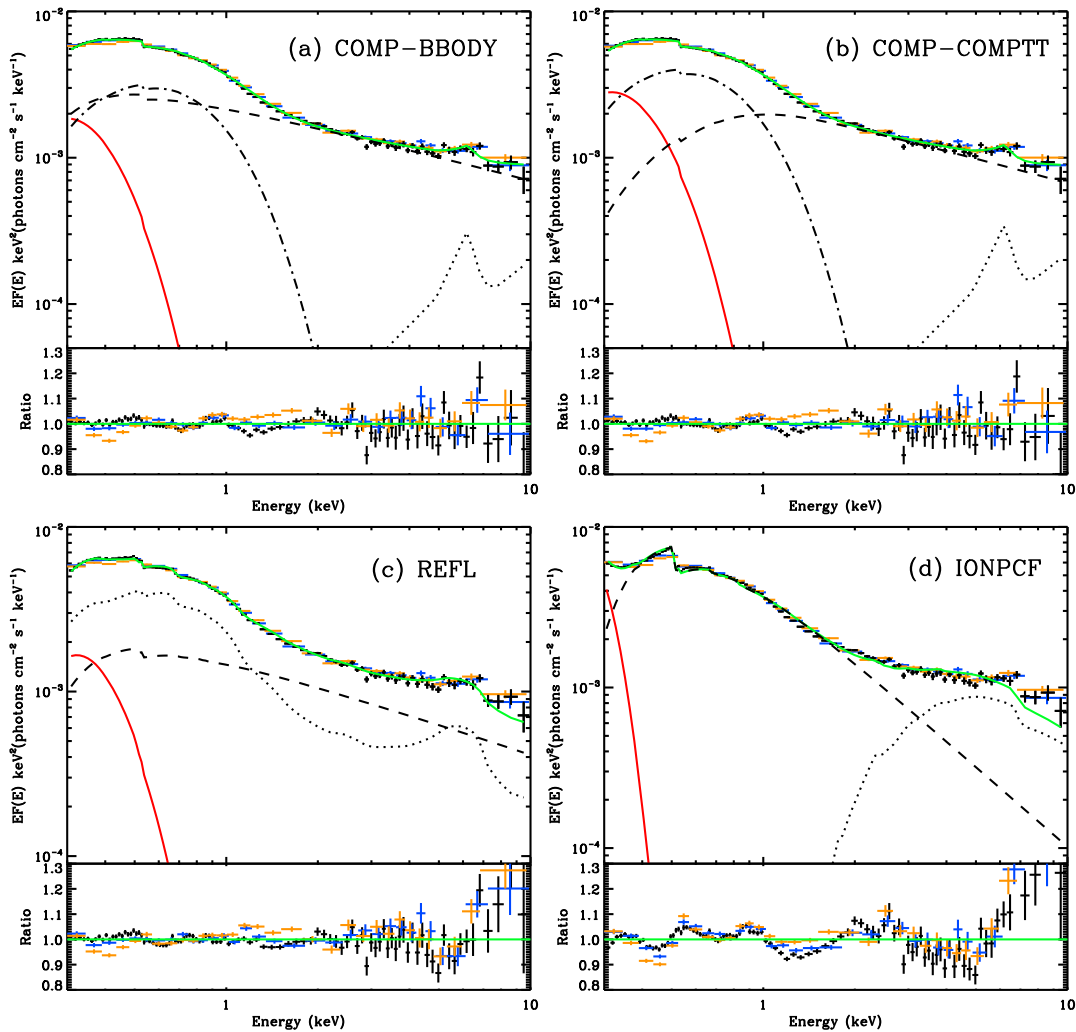}
\end{tabular}
\end{center}
\caption{Four methods of X-ray spectral fitting for PG 1244+026. In
  each panel, black, blue and orange points are the spectra from EPIC PN, MOS1, and
  MOS2 respectively. Model components are the blackbody (red solid
  line) and intrinsic coronal emission (black dashed line). In Panel-a
  and b, the dash-dotted line is the soft excess described by
  Comptonisation, while the dotted line is the neutral reflection. In
  Panel-a, the seed photons for the high temperature Comptonisation
  are from the blackbody component, while in Panel-b they are from the
  soft X-ray Comptonisation component.  In Panel-c, the dotted line is
  the smeared, ionised reflection component. In Panel-d, the dotted
  line is the fraction of power-law component after absorption by the
  partial-covering material. The data/model ratio is also plotted
  below each panel.}
\label{spec-fit-fig}
\end{figure*}

\section{Data Reduction}
\label{data-reduction-sec}

PG 1244+026 was observed for $\sim$123 ks with the {\it XMM-Newton}
satellite on the Christmas Day (Dec. 25th) 2011 (OBS ID:
0675320101). All three EPIC cameras were in small window mode to avoid
the photon pile-up effect. We used {\tt SAS} v12.0.1 and the latest
calibration files, and followed the standard procedures to reduce the
data. The entire observation period is free of high background flares,
therefore the resulting good exposure time is $\sim$86 ks for the PN
(due to the 71\% live time in PN small window mode) and $\sim$120 ks
for MOS (97.5\% live time in MOS small window mode). We chose the
source extraction region to be a circular region of radius 45\arcsec
for each EPIC camera. For the PN, the background region was chosen to
be a circle of radius 15{\arcsec} as far as possible from the source
while remaining on the same CCD chip. However, for the MOS cameras,
the background region was taken from a different CCD chip as the
central chip is fully occupied by the PSF of the source.  The total
source count rates are 11 ct s$^{-1}$, 2.2 ct s$^{-1}$ and 2.2 ct
s$^{-1}$ for PN, MOS1 and MOS2 respectively, which are all well below
the threshold count rates of causing photon pile-up effect in the
small window mode of each EPIC camera.

We selected data with {\tt{PATTERN} $\le$ 12} for MOS1 and MOS2, and
{\tt{PATTERN} $\le$ 4} for pn. Spectra were extracted from source and background
regions separately. The response matrices were produced using {\tt RMFGEN} and
{\tt ARFGEN}. The areas of source and background were calculated using {\tt
  BACKSCALE}. Source spectra were rebinned by {\tt GRPPHA} with a minimum of 25
counts per bin. All spectral fittings were performed in {\sc xspec} v12.7.1
(Arnaud 1996). Lightcurves were also extracted from both source and background
regions. Then the background lightcurve was subtracted from the source
lightcurve using {\tt LCMATH} in {\tt FTOOLS}. Note that the mean background
count rate within 0.3-10 keV is less than 0.3\% of source count rate for every
EPIC camera, and less than 20\% even at $\sim$ 10~keV.

We also used the pipeline RGS order spectrum to check for narrow
atomic features, grouped to a minimum of 10 counts/bin so that $\chi^2$
fitting is appropriate.

The OM UVW2, UVM2, UVW1, U, V band filters were used during the
observation period, with 5 exposures in every filter, each exposure
lasted 4600 seconds. We searched the merged OM source list file to
find the count rate in each filter associated with the source, and
pasted this into the OM data file template `om\_filter\_default.pi'
for use with the `canned' response files in spectral fitting (see
Section 6.2). We also extracted high resolution lightcurves from
individual exposure files, combined and rebinned to 100s (see Section
4).

\section{Time-averaged X-ray Spectral Modeling}

We assume that the 0.3-10~keV spectrum contains an intrinsic coronal
component which dominates the 2-10~keV bandpass.  We also assume that
there can be a contribution from the accretion disc itself, as the low
mass, high mass accretion rate of NLS1 such as PG 1244+026 means that
the disc can extend up to the soft X-ray bandpass (Done et al 2012:
hereafter D12).  We assume this disc emission above 0.3 keV
can be approximated as a blackbody
({\sc bbody} in {\sc xspec}).  The existence of a soft component close
to the X-ray bandpass means that the coronal emission is not likely to
remain as an unbroken power law at soft energies. If the disc or soft
X-ray excess is the source of seed photons then the downturn in the
Comptonised continuum is close to or within the observed soft X-ray
bandpass. We use the Comptonisation model {\sc nthcomp} in {\sc xspec}
(Zdziarski et al. 1996) so that this low energy turn-down is treated correctly. 

We then include additional model components in order to describe the
soft X-ray excess and remaining spectral features. We assume that all
components are absorbed by the Galactic column of $N_H=1.87\times
10^{20}$~cm$^{-2}$ (Kalberla et al. 2005), but include an additional
column of neutral absorption intrinsic to PG 1244+026 as a free
parameter ({\sc zwabs} with $z=0.048$). We assume an inclination of
$30^\circ$ for all reflection fits, as is probably appropriate for a
type 1 AGN. The total model is multiplied by a constant to account for
the slight difference in the normalization between the EPIC PN, MOS1
and MOS2 spectra (it was fixed at unity for the PN).

The models are discussed in detail below, with results plotted in
Figure~\ref{spec-fit-tab} and best-fit parameters listed in
Table~\ref{spec-fit-tab}.

\subsection{Comptonisation}

\subsubsection{Seed photons from the disc ({\sc comp-bbody})}

The blackbody and coronal emission alone give an unacceptable fit,
with $\chi^2_\nu=3969/1805$. Including a cool, optically thick
Comptonisation component ({\tt COMPTT} model in {\sc xspec}:
Titarchuk 1994) with seed photon temperature tied to the blackbody
temperature reduces this to $2312/1802$, showing clearly that the soft
excess is broader than a single blackbody. This continuum model has
$\Gamma=2.32\pm 0.02$. Adding neutral, relativistically smeared
reflection ({\sc kdblur*pexmon} Laor 1991, recoded as a convolution
model, Nandra et al. 2007) gives a significantly better fit
with $\chi^2_\nu=2261/1800$ for $\Omega/2\pi\sim 1.5$ for solar
abundance, and $R_{in}=16R_g$, steepening the intrinsic continuum to
$\Gamma=2.50_{-0.02}^{+0.04}$. 

The expression of the total model in {\sc xspec} is listed in
Table~\ref{spec-model-tab}, along with best-fit free parameters listed
in Table~\ref{spec-fit-tab}. The unfolded spectra are plotted in
Figure~\ref{spec-fit-fig}a.  The 2-10 keV photon index (hereafter:
$\Gamma_{2-10keV}$) is 2.50, which is among the steepest hard X-ray
slope in all NLS1s (e.g. Grupe 2004; Jin et al. 2012c). The
temperature for the low temperature Comptonisation is 0.16 keV, 
similar to the
$\sim$0.2 keV temperature seen for this component 
in all high mass accretion rate
AGN (Czerny et al. 2003; Gierli\'{n}ski \& Done 2004).  The blackbody
temperature is below the bandpass at 
60~eV but the Wien tail from this is significantly detected in the
data (removing the {\sc bbody} increases $\chi^2_\nu$ to $2361/1802$).

\subsubsection{Seed photons from Comptonisation ({\sc comp-comptt})}

We then assume that the seed photons for the coronal Comptonisation
component are
from the {\sc comptt} component, as appropriate for the geometry
sketched in D12 where the soft excess represents incomplete
thermalisation in the inner disc, and the corona is above the
inner disc. This mainly increases the norm of
both the blackbody and cool compton components as the higher seed
photon temperature reduces the low energy coronal flux.
We show this fit in Figure~\ref{spec-fit-fig}b, with best fit
parameters in Table~\ref{spec-fit-tab}.

The neutral reflection in both fits is rather larger than expected
for isotropic illumination. We caution that the lack of
signal-to-noise at high energies means that the preference for a
larger reflection fraction is probably driven by a preference for a
steep power law in the 0.5-5~keV bandpass, and that this in turn is
dependent on the detailed model assumed for the soft X-ray excess
(single electron temperature, single optical depth, single 
seed photon temperature and seed photons with a Wien spectrum). Better
high energy data is required to test how much reflection is really
present in these data. 

\subsubsection{Iron L emission line}

The models above are a good fit to the data, but there are significant
residuals at iron L, as seen previously (Fiore et al. 1998). Including
a Gaussian line gives $\chi^2_\nu=2129/1797$ (i.e. $\Delta\chi^2=132$
for 3 additional degrees of freedom) for a centroid energy of
$0.92$~keV with intrinsic width of 80~eV (so the line is too broad to
be seen in the RGS) and equivalent width of 12~eV. This line energy is
consistent with iron L emission, suggesting the presence of ionised
material, whereas the model here only includes neutral
reflection. Hence we investigate the ionised reflection models as the
origin of the soft X-ray excess.

\subsection{Ionised Reflection ({\sc refl})}

The fits above assumed that the soft excess was a true continuum
component, but an ionised reflector also includes a strong rise at low
energies due to the decrease in soft X-ray opacity of partially
ionised material and should also include features from iron L and
Oxygen emission (e.g. Fabian et al. 2009). We remove the {\sc comptt}
and {\sc pexmon} components and allow both the soft X-ray excess and
iron K$\alpha$ line to be fit by the same partially ionised reflector
with super-solar iron abundance (e.g. as seen in 1H 0707-495, Fabian et
al. 2009).  We model this using {\sc rfxconv}, which is based on the
ionised reflection tables of Ross \& Fabian (2005), recoded into a
convolution model (Kolehmainen, Done \& Diaz Trigo 2011). The
associated atomic features are not seen in the data, so the model
requires extreme relativistic smearing to match the observed, mostly
smooth continuum form. We follow previous work and allow the
emissivity to be a free parameter as well as $R_{in}$ (Crummy et al.
2006). This gives $\chi^2_\nu=2322/1802$, for an emissivity of $4$,
$R_{in}=3R_g$, a 2x overabundance of iron and ionisation parameter of
$\log \xi=3$, subtending a solid angle $\Omega/2\pi\sim 1.5$, where
the requirement for a large solid angle is mainly driven by the soft X-ray
excess. 

This reflection dominated model for the soft X-ray excess gives a very
different spectral decomposition around the iron K line than the {\sc
comp-bbody} and {\sc comp-comptt} models. Here there is a substantial
amount of partially ionised, strongly smeared reflection in the
4-10~keV spectrum, whereas before there was a smaller amount of
neutral, moderately smeared reflection.  However, the signal-to-noise
in the high energy bandpass is not sufficient to distinguish between
these two models.

Overall the fit is not as good as those derived above from
allowing an additional soft excess component and a free iron L line,
showing that a single ionised reflector is not able to simultaneously
describe the shape of the soft excess, iron K alpha line and iron L
line. Part of this may be because the low energy line emission in the
ionised reflection models is predominantly Oxygen rather than iron
L. However, the specifics of the line emission depend on the
underlying thermal structure of the reflector. The Ross \& Fabian
(2005) models assume that the reflector has no net flux entering from
below, whereas reflection from the disc would have an additional
ionisation level from collisional processes as well as
photo-ionisation.

\subsection{Partially ionised, partial covering absorption: {\sc ionpcf}}

Another scenario to model the X-ray spectrum is partial-covering
absorption by clumpy material along the line of sight (e.g. Miller et
al. 2007; Tatum et al. 2012). If this material is structured in a wind
then there can also be a contribution from scattering (e.g. Sim et al.
2010).  Such material could also be partially ionized since it is
directly illuminated by the central ionizing flux. We used {\sc xspec}
model {\tt zxipcf} (Reeves et al. 2008) to multiply the intrinsic
power law to produce the partially absorbed power law.

Adding a single partially ionised, partial covering component does not
describe the data well, with $\chi^{2}_{\nu}=3777/1802$, mainly due to
the failure of partial covering model
to follow the curvature of the soft excess around 1~keV. The
inferred power law index is extremely steep, with $\Gamma=3.6$, but
the only clear atomic features predicted by the model are 
some iron K alpha absorption lines around 6.6~keV. 

Including a second partial-covering component gives a better, but
still not acceptable fit, with  $\chi^{2}_{\nu}=3430/1801$
and the spectrum is still extremely steep, at $\Gamma=3.6$.

\begin{figure}
\includegraphics[scale=0.4]{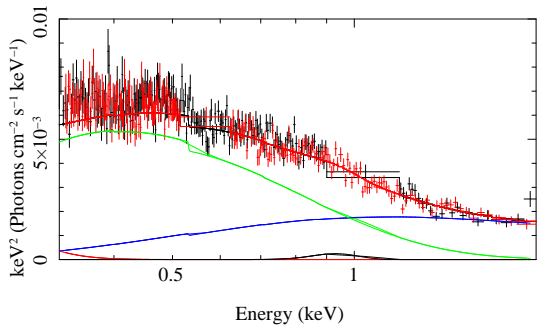}
\caption{The RGS1 (black) and 2 (red) spectra, devolvolved using the
  best fit {\sc comp-comptt} model ({\sc diskbb}: red line, {\sc
  comptt}: green line, {\sc nthcomp}: blue line, {\sc gaussian}: black
  line, see Section 3.1.3). This broad iron L line emission
  is consistent with the data, and there are no significant narrow absorption or
  emission features.}
\label{fig:rgs}
\end{figure}

%%%%%%%%%%%%%%%%%%%%%%%%%%%%%%%%%%%%%%%%%%%%%%%%%%%%%%%%%%%%%%%%%%%%%%%%%%%%%%%%%%%%%
\begin{table}
 \centering
   \caption{The best-fit value of the parameters in the four spectral fitting in
     Figure~\ref{spec-fit-fig}. The upper and lower limits give the 90\%
     confidence range.}
     \begin{tabular}{@{}llll@{}}
\hline
  Model & Component & Parameter & Value\\
\hline
{\tt COMP}   & \multicolumn{3}{l}{$\chi^{2}_{\nu}~=~2261/1800~=~1.26$} \\
{\tt :BBODY} & {\tt ZWABS}     & N$_{H}$ (10$^{22}$ $cm^{-2}$) & 0~$^{+0.001}_{-0}$\\
             & {\tt BBODY}     & kT (keV)                   & 0.058$\pm 0.002$\\
             & {\tt BBODY}     & norm                       & 1.3$\pm 0.1\times 10^{-4}$\\
             & {\tt NTHCOMP}   & $\Gamma$                   & 2.50~$^{+0.04}_{-0.02}$\\
             & {\tt NTHCOMP}   & kT$_{seed}$ (keV)          & tied\\
             & {\tt NTHCOMP}   & kT$_{e}$ (keV)             & 100 fixed\\
             & {\tt NTHCOMP}   & norm                       & 2.29$\pm 0.05\times 10^{-3}$ \\
             & {\tt COMPTT}    & kT$_{seed}$ (keV)          & tied\\
             & {\tt COMPTT}    & kT (keV)                   & 0.160~$\pm 0.003$\\
             & {\tt COMPTT}    & $\tau$                     & 96~$_{-60}^{+100}$\\
             & {\tt COMPTT}    & norm                       & 0.072~$_{-0.004}^{+0.06}$\\
             & {\tt KDBLUR}    & R$_{in}$ ($R_{g}$)           & 16~$^{+9}_{-6}$\\
             & {\tt PEXMON}    & Rel$_{refl}$                 & -1.5~$^{+0.3}_{-0.2}$\\
\hline
{\tt COMP}   & \multicolumn{3}{l}{$\chi^{2}_{\nu}~=~2263/1800~=~1.26$} \\
{\tt :COMPTT}& {\tt ZWABS}     & N$_{H}$ (10$^{22}$ $cm^{-2}$) & 0~$^{+0.001}_{-0}$\\
             & {\tt BBODY}     & kT (keV)                   & 0.062~$^{+0.001}_{-0.002}$\\
             & {\tt BBODY}     & norm                       & 1.8~$^{+0.1}_{-0.2}\times 10^{-4}$\\
             & {\tt NTHCOMP}   & $\Gamma$                   & 2.53~$^{+0.02}_{-0.03}$\\
             & {\tt NTHCOMP}   & kT$_{seed}$ (keV)          & tied to kT$_e$\\
             & {\tt NTHCOMP}   & norm                       & 2.1~$\pm 0.1 \times 10^{-3}$ \\
             & {\tt COMPTT}    & kT$_e$ (keV)              & 0.15~$^{+0.003}_{-0.003}$\\
             & {\tt COMPTT}    & $\tau$                     & 38~$^{+150}_{-20}$\\
             & {\tt COMPTT}    & norm                       &0.16~$\pm 0.01 $\\
             & {\tt KDBLUR}    & R$_{in}$ ($R_{g}$)           & 16~$^{+8}_{-6}$\\
             & {\tt PEXMON}    & Rel$_{refl}$                 & -1.76~$^{+0.32}_{-0.26}$\\
\hline
{\tt REFL}   & \multicolumn{3}{l}{$\chi^{2}_{\nu}~=~2322/1802~=~1.29$} \\
             & {\tt ZWABS}     & N$_{H}$ (10$^{22}$ $cm^{-2}$) & 0.015~$^{+0.004}_{-0.003}$\\
             & {\tt BBODY}     & kT (keV)                   & 0.053~$^{+0.004}_{-0.004}$\\
             & {\tt BBODY}     & norm                       & 1.3~$^{+2.3}_{-0.3}\times 10^{-4}$\\
             & {\tt NTHCOMP}   & $\Gamma$                   & 2.58~$^{+0.01}_{-0.01}$\\
             & {\tt NTHCOMP}   & norm                       & 1.61~$^{+0.09}_{-0.07}\times 10^{-3}$\\
             & {\tt KDBLUR}    & Index                      & 3.97~$^{+0.22}_{-0.13}$\\
             & {\tt KDBLUR}    & R$_{in}$ ($R_{g}$)           & 3.04~$^{+0.08}_{-0.02}$\\
             & {\tt RFXCONV}   & Rel$_{refl}$                 & -1.52~$^{+0.14}_{-0.56}$\\
             & {\tt RFXCONV}   & Fe$_{abund}$                 & 1.95~$^{+0.11}_{-0.09}$\\
             & {\tt RFXCONV}   & log$\xi$                   & 3.06~$^{+0.02}_{-0.02}$\\
\hline
{\tt IONPCF} & \multicolumn{3}{l}{$\chi^{2}_{\nu}~=~3430/1801~=~1.91$} \\
             & {\tt ZWABS}     & N$_{H}$ (10$^{22}$ $cm^{-2}$) & 0.088~$^{+0.002}_{-0.001}$\\
             & {\tt BBODY}     & kT (keV)                   & 0.050~fixed\\
             & {\tt BBODY}     & norm                       & $8.9\times 10^{-5}$~fixed\\
             & {\tt NTHCOMP}   & $\Gamma$                   & 3.61~$^{+0.01}_{-0.02}$\\
             & {\tt NTHCOMP}   & norm                       & 0.091~$^{+0.006}_{-0.010}$\\
             & {\tt ZXIPCF}    & N$_{H,1}$ (10$^{22}$ $cm^{-2}$) & 13~$^{+0.62}_{-0.39}$\\
             & {\tt ZXIPCF}    & log$\xi_{1}$                & 0.83~$^{+0.12}_{-0.14}$\\
             & {\tt ZXIPCF}    & {\it f}$_{cover,1}$          & 0.80~$^{+0.002}_{-0.011}$\\
\hline
   \end{tabular}
 \label{spec-fit-tab}
\end{table}
%%%%%%%%%%%%%%%%%%%%%%%%%%%%%%%%%%%%%%%%%%%%%%%%%%%%%%%%%%%%%%%%%%%%%%%%%%%%%%%%%%%%%%%

\subsection{High resolution spectra}

We fit the high resolution RGS data with the best fit {\sc
comp-comptt} model discussed in Section 3.1.2. All parameters were
frozen at the best fit values given in Table 2, but a constant offset
(best fit value of 0.92) was allowed, to take into account possible
calibration differences between the gratings and CCDs. This gives a
good fit, with $\chi^2_\nu=4268/4205$.  We then include the iron L
line, again fixed that seen in the CCD detectors, and find
$\chi^2_\nu=4256/4205$, so this weak, broad feature is consistent with
the grating spectra. Figure~\ref{fig:rgs} shows the RGS data
de-convolved with this model. It is clear that there are no narrow
absorption features evident by eye. We quantify this fitting a warm
absorber {\sc zxipcf}, assumed to cover the entire source. We allow
this to be outflowing to a maximum velocity of $0.1$~c. This model is
only tabulated down to columns of $5\times 10^{20}$~cm$^{-2}$, but the
fit pegs to this value, giving $\Delta\chi^2$ of $5$ {\em worse} than
no absorption at all for $\log\xi=3$. The fit becomes progressively
worse at lower ionisations, with $\Delta\chi^2$ increased to  $15$ at
$\log\xi=2.4$ and $110$ for $\log\xi=1.6$.

Thus it is clear that none of the curvature below 2~keV seen in the CCD spectra
is due a classic `warm absorber', though we note that the
{\sc ionpcf} fits in Section 3.3 (which model the complexity around the
iron K$\alpha$ line by partial covering) do not predict any features below
2~keV.

\subsection{Summary of spectral fits}

The main question addressed in this paper is the origin of the soft
X-ray excess, whether it is predominantly a separate soft continuum
component, or extremely smeared reflection or due to the spectral
curvature associated with complex absorption. These models are not
necessarily mutually exclusive, nor are they the only models which can
be envisaged. For example, we used neutral reflection in the models
where the soft excess was fit by a separate components ({\sc
comp-bbody} and {\sc comp-comptt}).  The reflector could instead be
ionised, giving some contribution to the soft X-ray excess, while
there may be an additional true soft continuum which contributes to
the {\sc refl} or {\sc ionpcf} fit. There could also be multiple
reflectors (e.g Fabian et al 2002) or absorbers (e.g. Miller et al
2007). We do not explore these composite models further 
as we are
concerned with the origin of the majority of the soft X-ray
excess. The spectral fits alone rule out complex absorption as a major
contributor to the soft X-ray excess, so in what follows we use
variability to try to distinguish between a separate soft component
and an extremely smeared reflection origin for the majority of the
soft X-ray excess.

\section{Lightcurve and Variability}
\label{var-sec}

\begin{figure}
\begin{center}
\includegraphics[clip=,scale=0.67]{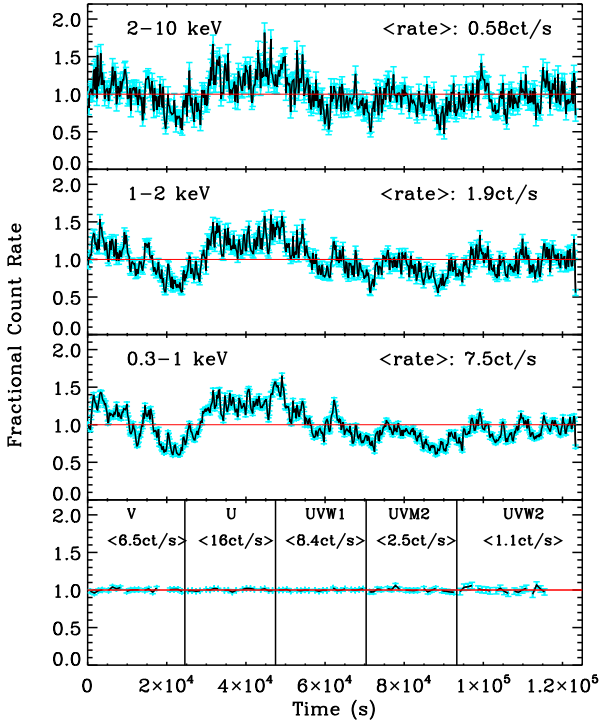}
\end{center}
\caption{The 100s binned light-curves in each X-ray energy band and the five OM filters, divided by the mean count-rates given in the bracket. Error-bars are shown in cyan.}
\label{lc-fig}
\end{figure}

%-----------------------------Figure Start------------------------------
\begin{figure*}
\centering
\includegraphics[angle=90,scale=0.6]{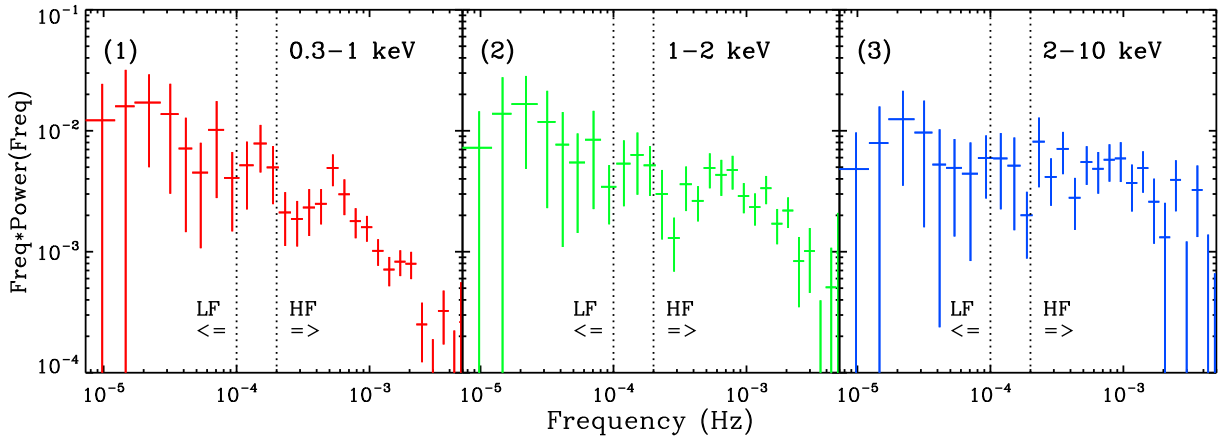}
\caption{The power spectrum of each energy band after subtracting the Poisson
  counting noise, calculated by {\tt FTOOL} task {\tt POWSPEC}, rebinned as a
  geometrical series with geometrical step equal to 1.2.}
\label{powspec-fig}
\end{figure*}
%-----------------------------Figure End--------------------------------

\subsection{Power spectra}

Figure~\ref{lc-fig} shows the light-curves in both soft, medium and
hard X-ray, together with the simultaneous optical/UV data from the
OM. These is no significant variability in the optical/UV but the
X-rays show fluctuations on all time-scales from tens to thousands of
second, with subtle changes in the amount of
variability with energy. We study this in more detail by taking the
power spectral density (PSD).

Each X-ray light-curve shown in Figure~\ref{lc-fig} was a combination of
three 100s-binned background subtracted light-curves from the three
EPIC cameras separately. The {\tt FTOOL} task {\tt LCMATH} was used to
perform this light-curve merging. Then {\tt FTOOL} task {\tt POWSPEC}
was used to perform a DFT on each of these three light-curves to derive
the periodogram, which is a realisation of the intrinsic continuous PSD
(e.g. Vaughan et al. 2003). The norm parameter of {\tt POWSPEC} was set to
$-2$, such that white noise is subtracted and the remaining power
integrates to give the excess variance in the lightcurve. These PSDs
are plotted as $fP(f)$ in Figure~\ref{powspec-fig}. 

These show differences between the PSD shapes at each energy, with the
2-10~keV bandpass showing more power at the highest frequencies than
the lowest energy band. This is a common feature of NLS1 power spectra
(e.g. M$^{c}$Hardy et al. 2004). There is also some potential structure in
the PSD of the 0.3-1~keV lightcurve (left panel in
Figure~\ref{powspec-fig}) around ${\sim}5{\times}10^{-3}$Hz, possibly
resembling the double Lorentzian fit to the PSD of Ark 564 (M$^{c}$Hardy et
al. 2007) though the statistical significance of this would require
widespread simulations to assess, which are beyond the scope of this
paper.

\subsection{Frequency-dependent Fractional {\it RMS} Spectra}

We have divided the 0.3-10 keV into different energy bands and
explored their PSDs. But we can also divide the observed frequency
range into bands, and explore the more detailed energy-dependence of
the fractional variability for each frequency band, i.e. the {\it RMS}
spectrum (e.g. Edelson et al. 2002; Markowitz, Edelson \& Vaughan
2003; Vaughan et al. 2003).

There are two ways to calculate the {\it RMS} spectra in the
literature. The most straightforward is where the light
curve (length $T$, bin time $\Delta t$) is divided into $M$
segments. The excess variance in each segment is averaged together to
give a measure of the {\it RMS} over a frequency range from $M/T$ to $1/(2\Delta
t)$, with error given in Vaughan et al (2003). However, red noise leak
can be an issue with this if there is substantial variability power at
frequencies lower than $M/T$. We instead use the alternative approach,
which calculates the {\it RMS} by integrating the power spectrum of the
full lightcurve (from $1/T$ to $1/(2\Delta t)$) 
only over the frequency range of interest, as this
suppresses red noise leak. We calculate errors following Poutanen,
Zdziarski \& Ibragimov (2008).  This uses Poisson errors, so non-gaussianity of 
low counts is not an issue at high energy, but it is not clear how to treat a zero 
count bin. We choose the smallest energy bands
which are compatible with not having more than 10\% of the bins with zero counts.

We also tested whether the background subtraction would affect our
calculation, especially at high energies above 4 keV where the count
rate is low. We found the background count rate was only 10-20\% of
the source count rate in the 8-10~keV bin, and much less at lower
energies. The background subtracted source {\it RMS} in the HF band is
0.13$\pm$0.02, not significantly different to the non-background
subtracted HF {\it RMS} in this band of 0.14$\pm$0.02. Therefore, we
conclude that our results were not affected by background subtraction.

\begin{figure}
\begin{center}
\includegraphics[clip=,scale=0.5]{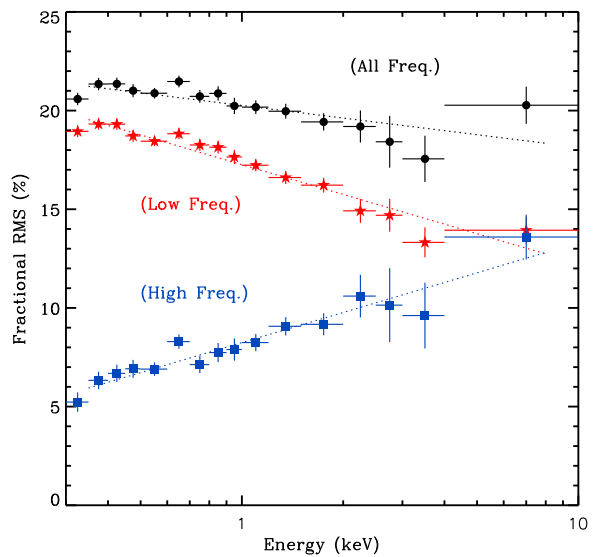}
\end{center}
\caption{The fractional {\it RMS} spectrum for each frequency band. Dotted lines
are simple linear fitting to every spectrum to highlight the spectral trend.}
\label{frac-rms-tot-fig}
\end{figure}

The fractional {\it RMS} spectrum for the whole observed frequency band
i.e. from $8\times 10^{-6}-5\times 10^{-3}$~Hz (123ks-200s) is shown as the
black points in Figure~\ref{frac-rms-tot-fig}. The total {\it RMS} spectrum is
relatively flat, with a small dip at 3$\sim$4 keV. This is an unusual shape, as
it is not similar to the 1-2~keV peaked shape seen in the extreme low state
NLS1, which may indicate that the spectra are dominated by reflection
(e.g. Fabian et al 2004), nor is it similar to the high state NLS1 (e.g. RE
J1034+396) where the low energy variability is strongly suppressed, which may
indicate that the soft excess is a separate component (Middleton et al. 2009;
Jin et al. 2009).

However, RE J1034+396 showed different {\it RMS} at different frequencies
(Middleton et al. 2009). We explore whether the same is true for PG 1244+026, by
dividing the data into two frequency bands, i.e. low frequency
([$8\times10^{-6}$Hz, $1\times10^{-4}$Hz] or [10ks, 123ks], hereafter: LF, red
points), and high frequency ([$2\times10^{-4}$Hz, $5\times10^{-3}$Hz] or [200s,
  5ks], hereafter: HF, blue points). The {\it RMS} spectra for these bands are
shown in Figure~\ref{frac-rms-tot-fig}. It is clear that at LF, the soft X-rays
are more variable than the hard X-rays; while the opposite is true at HF.  This
clearly shows the reason for the unusual total {\it RMS} spectrum, which is
because it is the sum of two very different spectral behaviours at high and low
frequencies.

\begin{figure*}
\begin{tabular}{cc}
\includegraphics[scale=0.52]{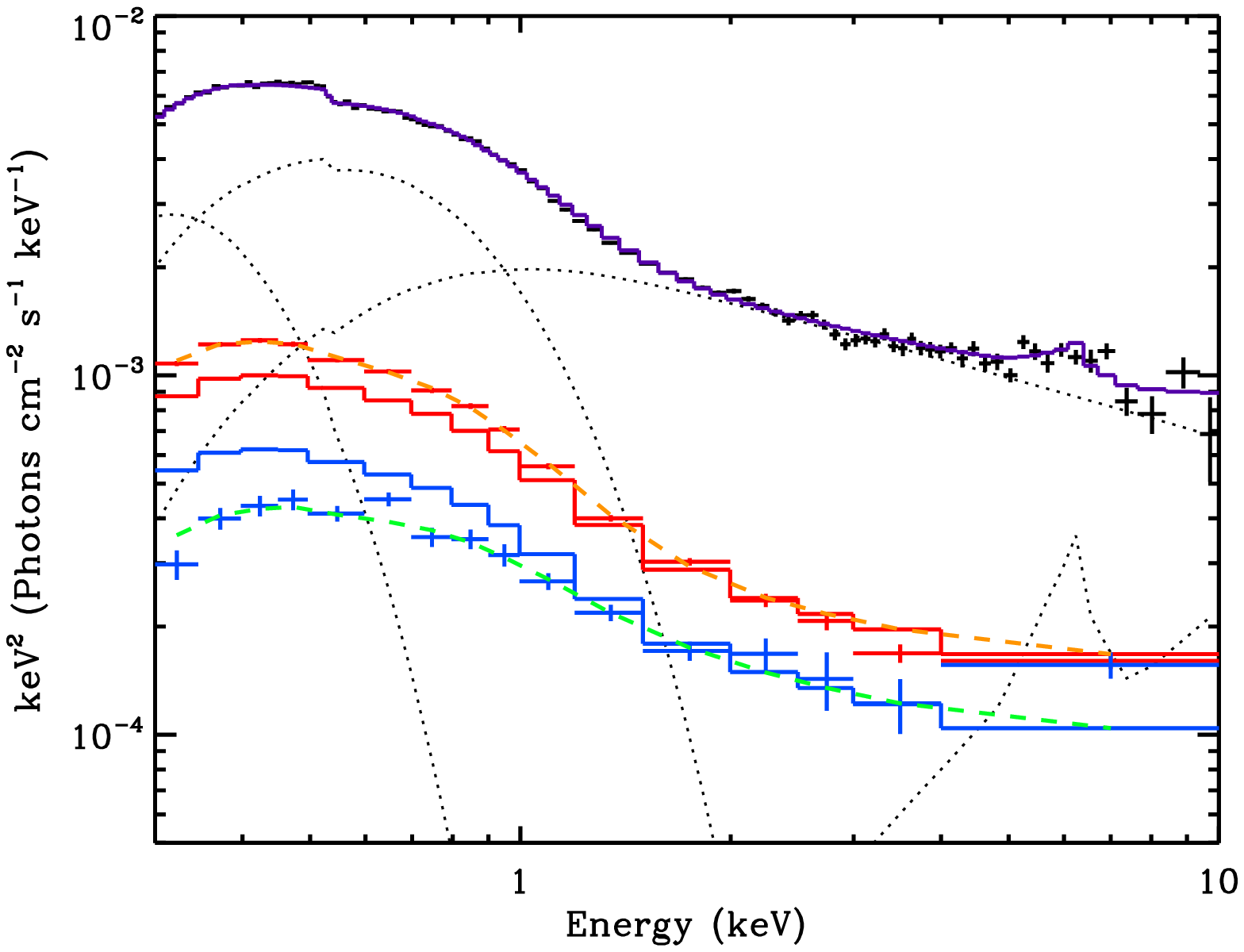} &
\includegraphics[scale=0.45]{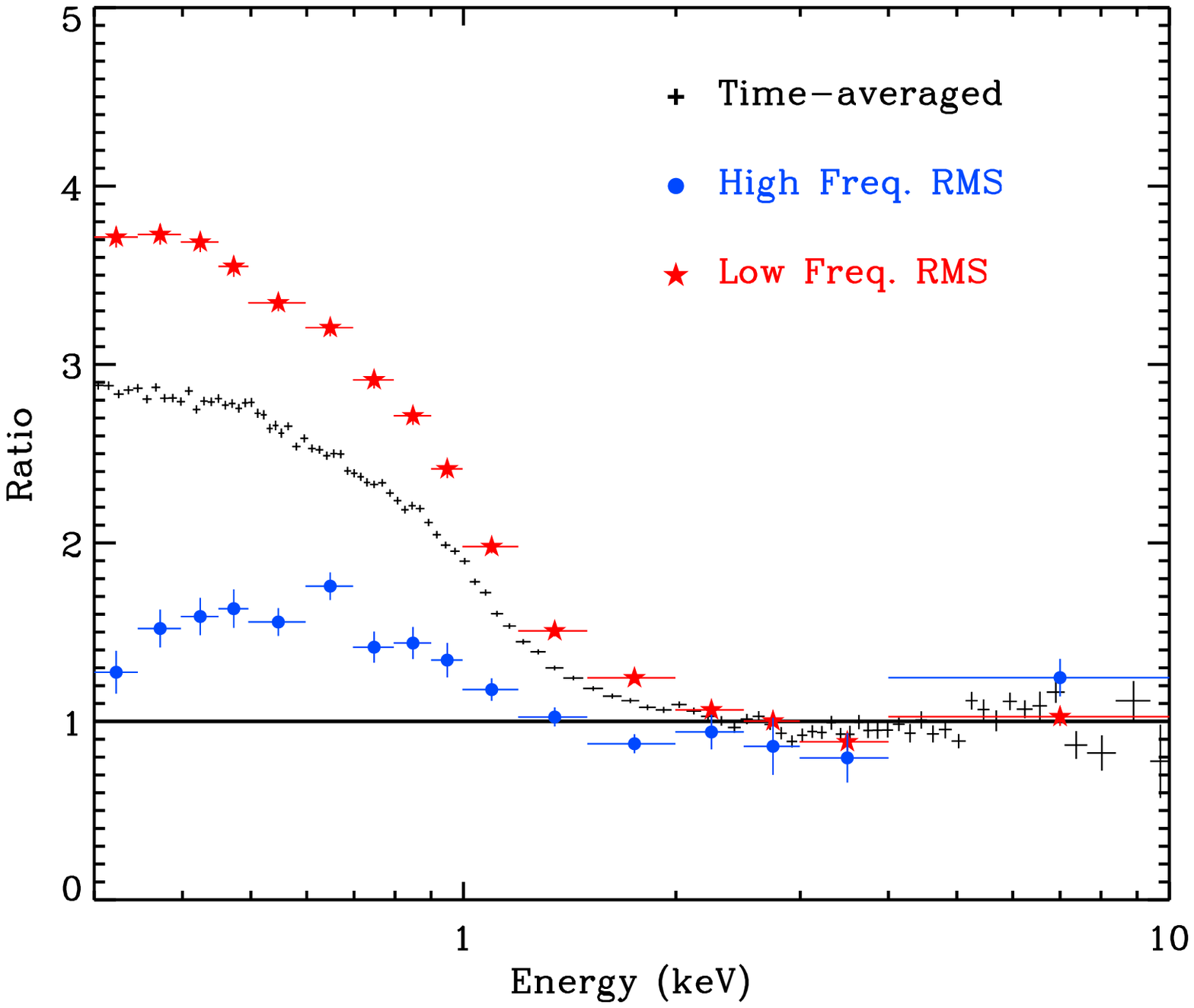}\\
\end{tabular}
\caption{a) Comparison between the time-average spectrum (black) and
{\it RMS} spectra for HF (blue) and LF (red), respectively. The black
lines show the best fit model components to the time averaged
spectrum, with blackbody, cool, optically thick Comptonisation, hot,
optically thin Comptonisation and neutral reflection shown as the
dotted lines. The blue and red solid lines show the total model scaled
by 0.1 and 0.15 respectively to match the HF and LF power law. The
dashed blue and red lines show the model assuming both blackbody and
soft excess are further scaled by a factor 0.6 and 1.3 for HF and LF
respectively. b) Ratio of each spectrum to its best fit 2-10~keV power
law highlighting the changing strength of the soft X-ray excess with
variability timescale.}
\label{abs-rms-spec}
\end{figure*}

\subsection{Frequency-dependent Absolute {\it RMS} Spectra}

We explore the frequency dependent spectral behaviour in more detail
by multiplying the fractional {\it RMS} in each energy bin by the mean
count rate in that bin to produce absolute {\it RMS} spectra. These
can then be fit directly in {\sc xspec} using the standard instrument
response (e.g. Revnivtsev et al. 2006; Sobolewska \& \.{Z}ycki 2006;
Middleton et al. 2011). PN and MOS produced identical lightcurves, but
they have different response files and slightly different spectral
normalisation. Therefore we only used the mean count rate from the PN
time-averaged spectrum to calculate the absolute {\it RMS}, and used
the PN response file in {\sc xspec} fitting.

We compare the absolute {\it RMS} spectra with the PN time-averaged
spectrum in Figure~\ref{abs-rms-spec}a using the {\sc comp-comptt}
model from Section 3.1.2, multiplied by 0.1 and 0.15 (for HF and LF,
respectively). Plainly, this over-predicts the soft X-ray variability
in the HF and under-predicts it in the LF. Figure~\ref{abs-rms-spec}b
illustrates this dependence of the soft X-ray excess on variability
timescale by showing the ratio of each spectrum with the best fit
2-10~keV power law. The soft X-ray excess is much stronger relative to
the power law emission in the low frequency variability spectrum than
in the fast variability spectrum. This is as expected from the power
spectral results (Figure~\ref{powspec-fig}), as these show that there
is more variability in the hard band than the soft at high
frequencies, and more variability in the soft band than the hard at
low frequencies. The dashed lines in Figure~\ref{abs-rms-spec}a show
the model assuming that the blackbody and soft excess are scaled by a
factor 1.3 relative to the coronal emission in the LF (red dashed),
and a factor 0.6 in the HF (blue dashed).  This models the variability
well, though the HF {\it RMS} are slightly over-predicted at the
softest energies, perhaps indicating that the blackbody is not varying
as strongly as the soft X-ray excess.

The alternative {\sc refl} models (section 3.2) can also match these
data, this time by changing the amount of ionised reflection (to
$\Omega/2\pi\sim 2.5$ and $1$ for LF and HF respectively, compared to
$1.5$ in the time averaged spectrum. These models give a slightly
better match to the rise in the final HF {\it RMS} spectral bin
between 4-10~keV (as discussed in Section 5.1), though we caution that
the lack of statistics (in particular the number of bins with zero
counts) may be an issue here.

\begin{figure*}
\begin{tabular}{cc}
\includegraphics[scale=0.52]{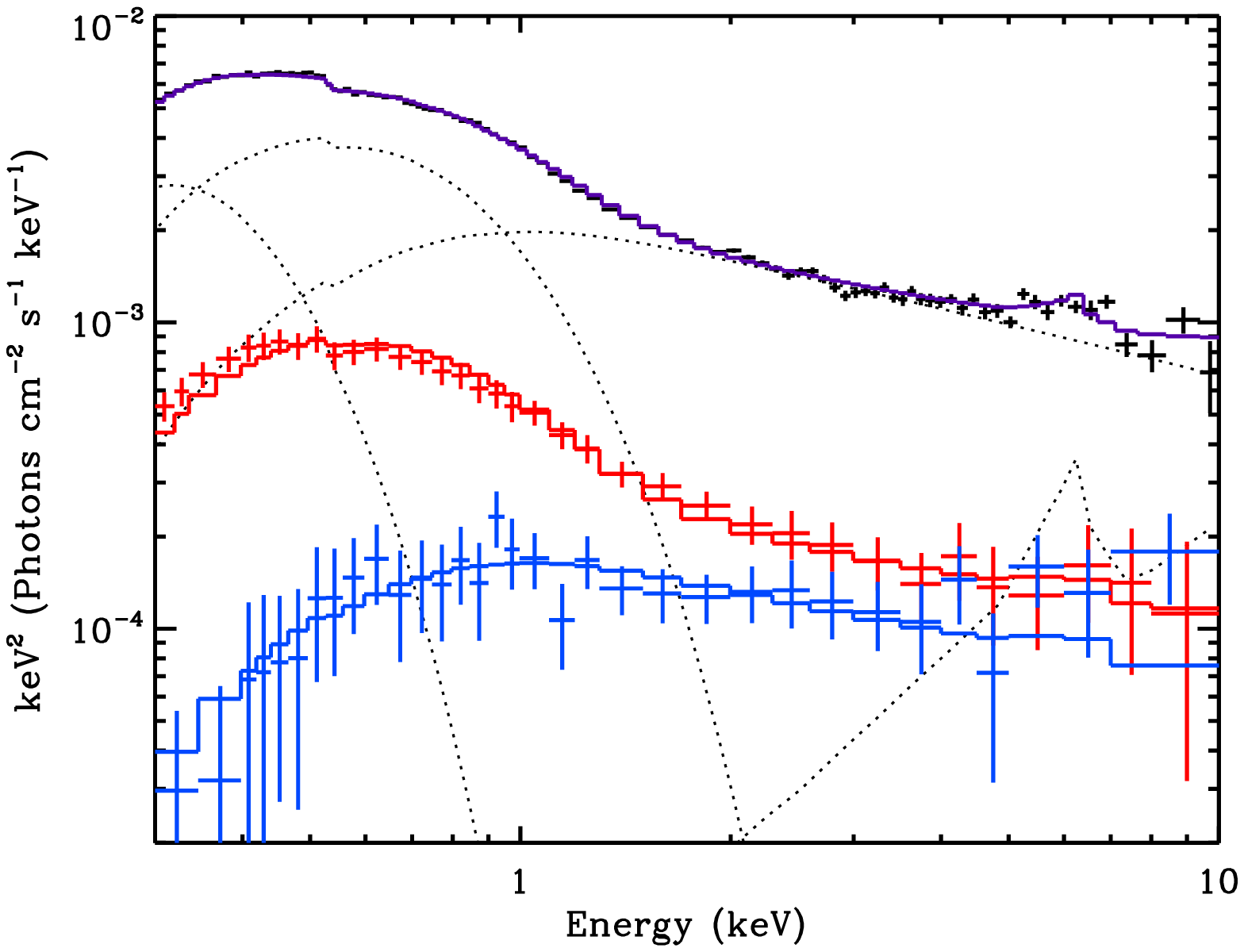} &
\includegraphics[scale=0.45]{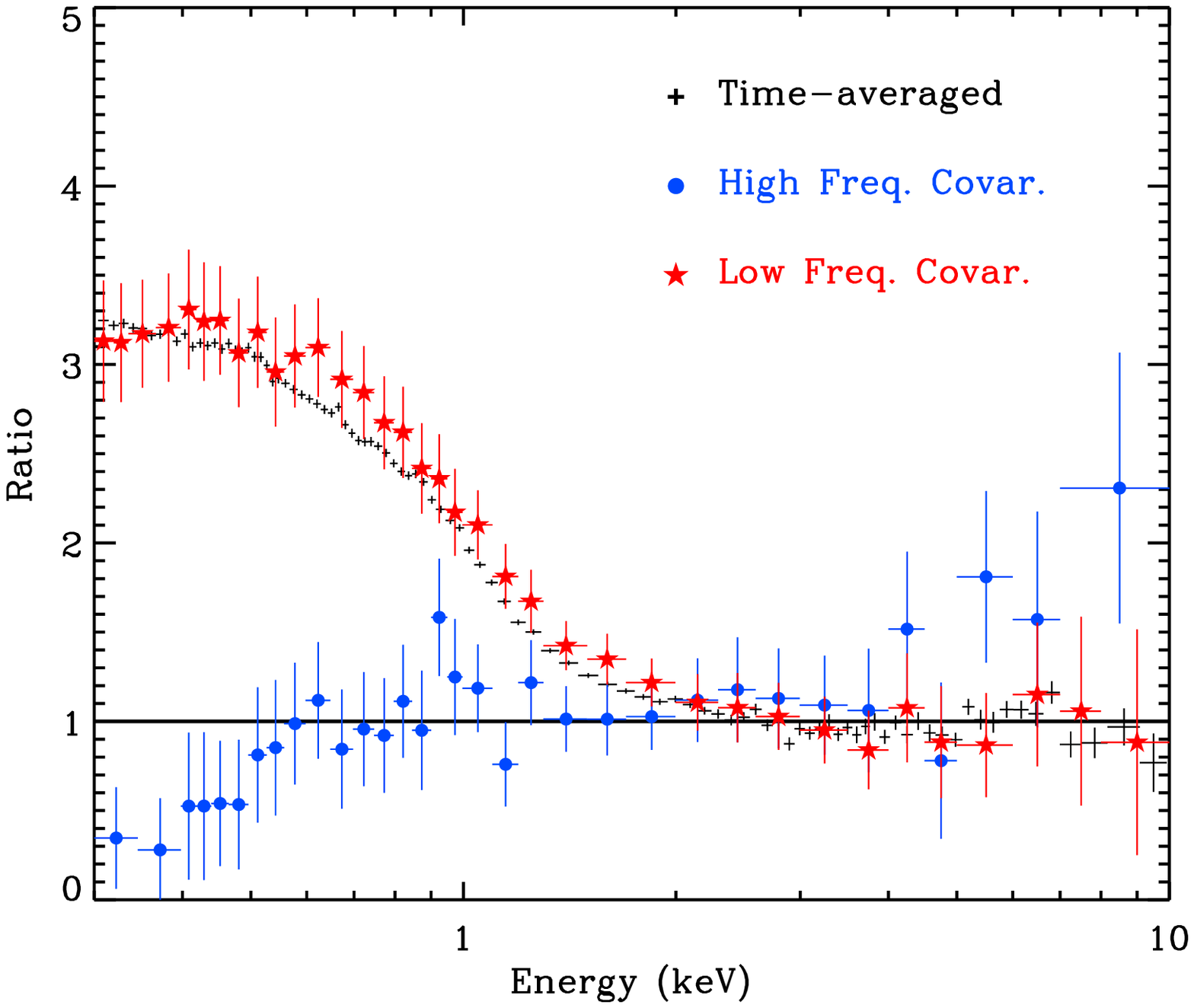}
\end{tabular}
\caption{a) As for Figure~\ref{abs-rms-spec} a and b, but with the
4-10~keV covariance spectra for HF (blue) and LF (red), respectively.
The 4-10~keV HF covariance spectrum clearly shows no component of the soft
X-ray excess or soft blackbody emission.}
\label{abs-cov-spec}
\end{figure*}

\section{Frequency-dependent Covariance Spectra}

The {\it RMS} spectrum shows the total variability at a specific
frequency range in every energy bin, but the very limited signal-to-noise at high energies
prevented us from reducing the width of energy bin further in the {\it RMS} spectrum.
One way to increase the spectral resolution of variability is to look only for
the correlated variability via the `covariance spectrum', a technique
developed by Wilkinson \& Uttley (2009). A `reference band' can
be chosen to give a high signal-to-noise lightcurve and
cross-correlated with the lightcurve in each individual band. The
correlated variability in each energy band has much smaller error bars
as it removes the uncorrelated white noise variance. This makes it a more
sensitive technique so it can be used to explore the variability at
higher energies. It also has obvious advantages in that it
explicitly pulls out the correlated variability. Both the power law
and its reflection contribute to the 0.3-1~keV and 4-10~keV bands if
reflection makes the soft X-ray excess,
whereas these energies are connected only by the power law in the
Comptonisation models.

We derive the covariance spectrum for the HF and LF frequency ranges again
using the power spectra.  We used a DFT to transform the light-curve
into a periodogram; then kept a specific frequency range while letting
all the power outside this range be zero; then used the inverse DFT to
transform the band-limited periodogram back to a new light-curve. This
new light-curve has the same mean count-rate as the original
light-curve, but only contains variability over the chosen
frequency range. We applied the frequency filter to the light-curve in
every energy bin, and then followed the procedure in Wilkinson \&
Uttley (2009) to derive the covariance spectrum without dividing the
light-curve into small segments. The resultant covariance spectrum is
only for the specific frequency range, and so is frequency-dependent.
When the energy bin for which the covariance is being calculated is
within the reference band, we recalculate the reference band
lightcurve excluding that channel so that Poisson errors are never
included in the correlated signal.

This should also pull out any correlated variability which has a short
(with respect to the binning) time lag/lead. Recent studies of the similar
mass and mass accretion rate object 1H 0707-495 have shown a lag of 30s
of the soft band behind the high energy variation at high frequencies (Fabian et al.
2009). This is within our bin time of 100s so we would include any
component which lags or leads on this timescale in our covariance
spectrum.

\begin{figure}
\begin{center}
\includegraphics[scale=0.52]{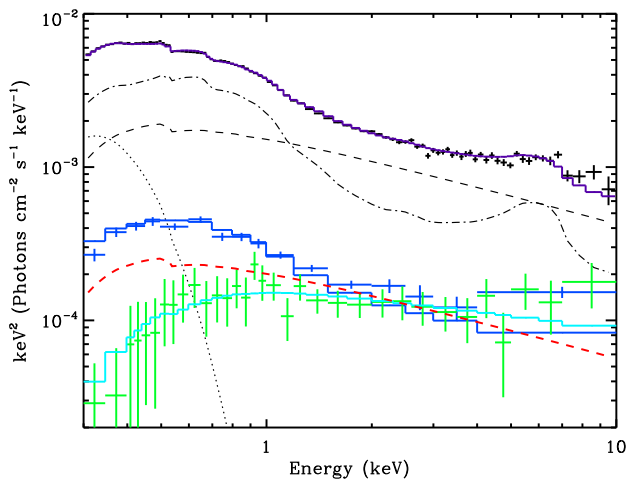}
\end{center}
\caption{Comparison between the time-average spectrum (black), HF {\it
    RMS}
spectrum (blue) and the HF 4-10~keV covariance spectrum (green). The
time-averaged spectrum is shown decomposed with the {\sc refl}
model. The HF {\it RMS} spectrum does show variability in the soft X-ray
excess but the lack of this in the HF covariance component shows that
none of this fast variability in the soft excess can correlate with
the 4-10~keV flux changes, ruling out the  {\sc refl} model. 
Furthermore, the turndown in the covariance spectrum below 1~keV
clearly shows that the seed photons for the continuum are at soft
excess energies rather than from the disc component as assumed
in the {\sc refl} model (dashed red line).}
\label{newfig_rms_he_cov}
\end{figure}

\subsection{High Energy Covariance: 4-10~keV reference band}

We choose 4-10 keV as the reference band, as this is the one where the
Comptonisation and reflection models for the soft X-ray excess show
most difference. This band contains both the soft excess component
(ionised, blurred reflection) and the continuum in the {\sc refl}
model, but does not contain any of the soft X-ray excess in the {\sc
comp-comptt} model.

Figure~\ref{abs-cov-spec}(a) shows the LF (red) and HF (blue)
covariance spectra, compared to the time averaged spectrum, while (b)
shows the ratio of each spectrum to its best fit 2-10~keV power
law. The LF covariance spectrum is very similar to the time averaged
spectrum, but the HF covariance is completely different, lacking all
soft excess and blackbody components. Thus on timescales of 5000-200s
{\em none} of the soft excess variability seen in the HF {\it RMS}
spectrum correlates with the hard X-ray lightcurve.

We show this in more detail by showing the HF {\it RMS} (blue) and
covariance (green) spectra together with the time averaged spectrum
(black) in Figure~\ref{newfig_rms_he_cov}. There is clearly more soft
variability in the {\it RMS} spectrum than in the covariance
spectrum. Thus there is some fast variability in the soft excess (as
shown by the power spectra), but this is uncorrelated with the fast
variability in hard X-rays (or correlated with a lag/lead much longer than
the 100s bin time). The black lines show the best fit reflection model
to the time averaged spectrum. The blue solid line shows the best fit
model of these components to the HF {\it RMS} spectrum, resulting in
the blackbody normalisation going to zero, while the amount of
reflection reduces to $\Omega/\pi \sim 1$ (as discussed in Section
4.3). The red dashed line shows a similar fit to the HF covariance
spectrum (green). Even with the blackbody and reflection
normalisations set to zero, the intrinsic power law in the reflection
fit is too steep to match the correlated spectrum.  The cyan solid
line shows instead the coronal `power law' emission from the
Comptonisation model for the soft X-ray excess ({\sc comp-comptt}).
Plainly this is a much better fit to the slope of the covariance
spectrum, as well as having the downturn for seed photons at the
correct energy assuming that these are from the soft excess.

\begin{figure*}
\begin{tabular}{cc}
\includegraphics[scale=0.52]{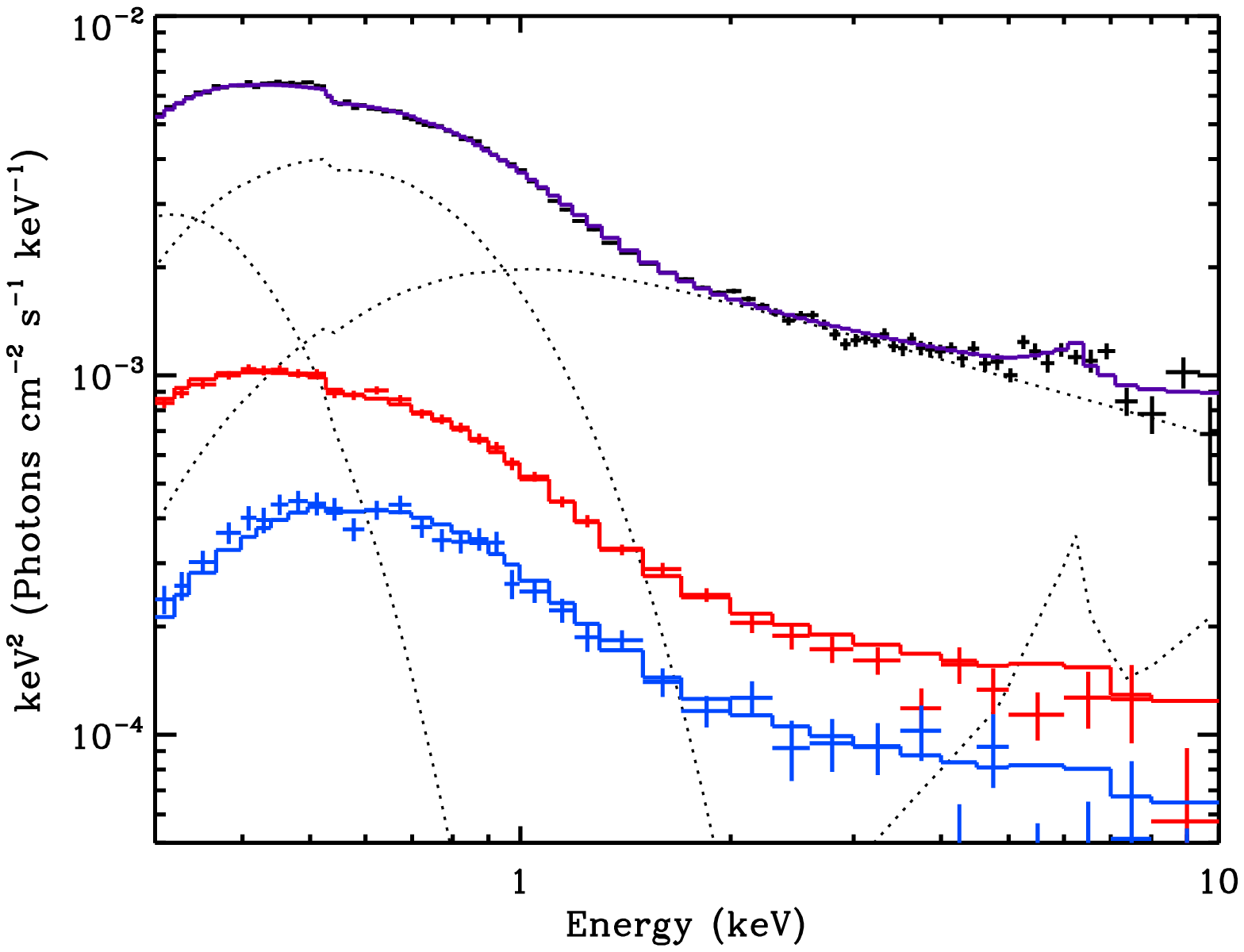} &
\includegraphics[scale=0.45]{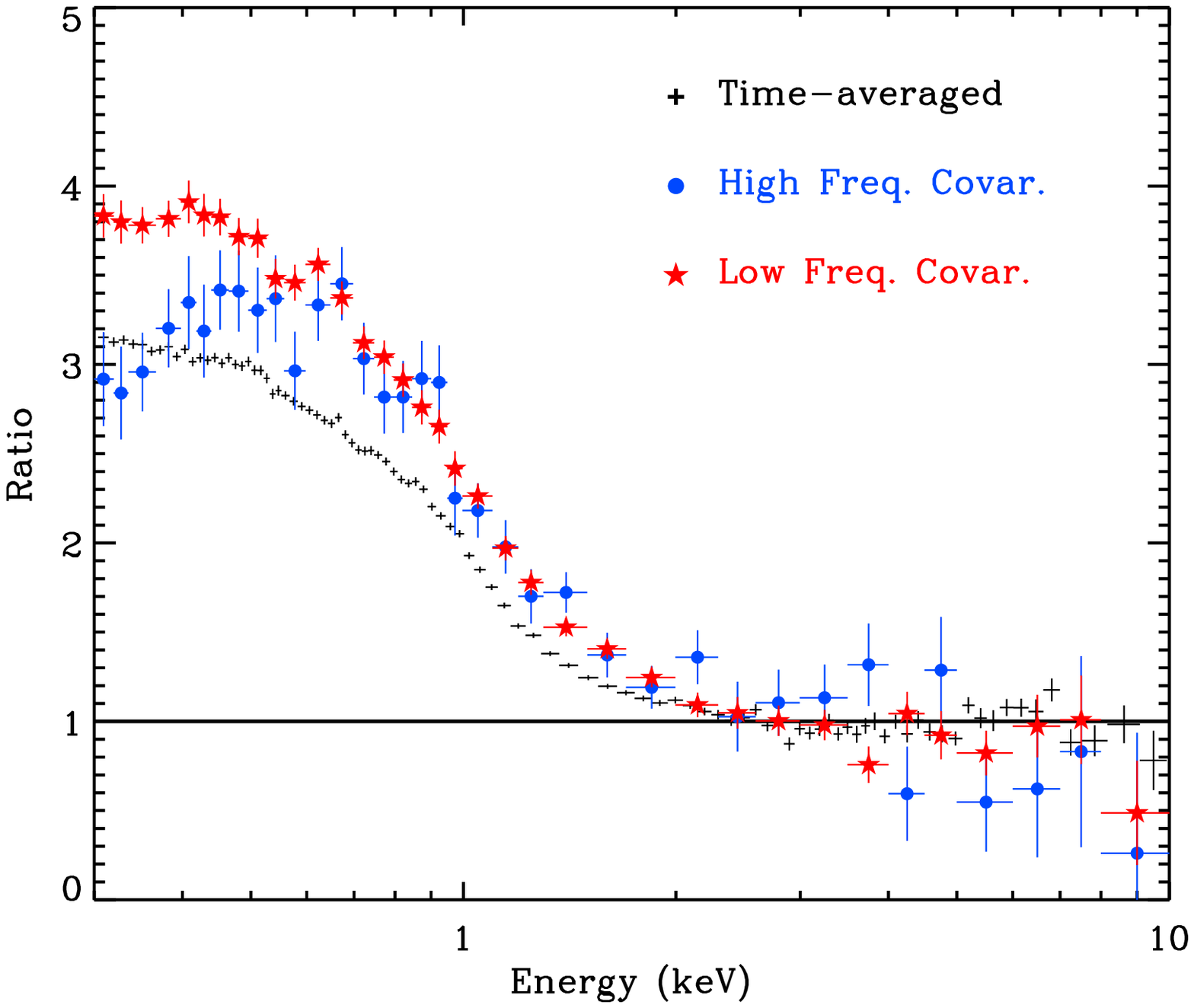}
\end{tabular}
\caption{As in Fig~\ref{abs-rms-spec}a and b but using 0.3-1 keV
as the reference energy band. a) The blue and red solid lines show the
total model scaled by 0.08 and 0.16 respectively to match the HF and
LF power law, except that the hot, optically thin Comptonisation was
further scaled down by 0.85 in both HF and LF models, and blackbody
was removed from the HF model.  b) the ratio of the spectra to their
2-10~keV best fit power law. The LF and HF components are very similar
except for the drop in HF covariance at the lowest energies,
indicating the physical reality of the additional blackbody
component.}
\label{abs-lecov-spec}
\end{figure*}

The key issue is the difference between the {\it RMS} and covariance
spectra shows that there is variability in the soft X-ray excess which
is not correlated with variability in the 4-10~keV bandpass. Yet in
reflection models, the soft X-ray excess variability contributes to
variability in the 4-10~keV bandpass so these cannot be uncorrelated.
Another difference is that the intrinsic `power law spectrum' in the
reflection model is significantly steeper than in the Comptonisation
model. Yet the covariance spectrum alone fit to a hot (temperature
fixed at 100~keV) Comptonisation component gives $\Gamma=2.25\pm 0.2$,
significantly flatter than the $\Gamma=2.58\pm 0.01$ derived from the
time averaged spectrum reflection fit, but consistent with the overall
(including reflection) spectral index of $\Gamma=2.32\pm 0.02$ seen in
the Comptonisation model. The seed photon energy of $0.13\pm
0.04$~keV is also consistent with the soft excess being the source of
seed photons for the hot corona, and not with the lower temperature
disc.

\begin{figure}
\begin{center}
\includegraphics[bb= 79 10 564 700, clip=1,scale=0.34,angle=270]{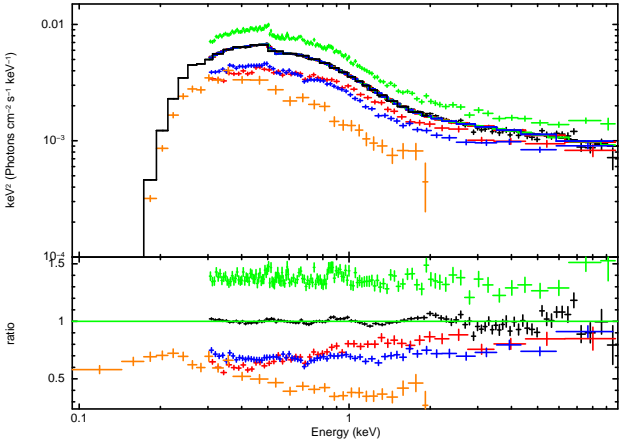}
\end{center}
\caption{The short and long term X-ray variability. The black spectrum is the
  time-averaged {\it XMM-Newton} EPIC PN spectrum from the 2011 observation,
  with the blue and green spectra extracted from all time intervals of count
  rates ${\le}$0.75 and ${\ge}$1.25 of the mean count rate separately. The red
  spectrum is the time-averaged {\it XMM-Newton} EPIC PN data from the 2001
  observation. The orange spectrum is from the 1991 {\it ROSAT} PSPCB observation.
  The lower panel shows the ratio between these spectra and the
  best-fit Comptonisation model to the black spectrum.}
\label{flux-var-fig}
\end{figure}

\subsection{Low energy covariance: 0.3-1~keV reference band}

The difference between the HF {\it RMS} and 4-10~keV HF covariance
spectra shows that the soft excess varies independently from the high
energy continuum. Figure~\ref{abs-lecov-spec}(a) shows the HF and LF
covariance spectra extracted with a reference band 0.3-1~keV to focus
on this variability, while (b) shows these as a ratio to the (scaled)
high energy power law emission. The difference between the HF and LF
covariance spectra is much less dramatic using the low energy
lightcurve as the reference band. The only significant difference
between the LF and HF 0.3-1~keV covariance spectra are that the HF
covariance spectrum dips at the lowest energies, indicating that there
is a separate component, described here as an additional blackbody,
which contributes only below 0.5~keV.

We fit the covariance spectra using the {\sc comp-comptt} model.  Fig
\ref{abs-lecov-spec}(a) shows this best fit time average model scaled
by 0.16 (LF, red) and 0.08 (HF, blue), respectively, but with the
blackbody component removed from the HF model. This matches the
downturn seen in the HF covariance spectrum at the lowest energies,
showing clearly that the soft X-rays are composed of two components,
in addition to the extrapolated high energy emission. This softest
emission component is most plausibly the disc for this low mass, high
mass accretion rate NLS1 (see section 6).

%%%%%%%%%%%%%%%%%%%%%%%%%%%%%%%%%%%%%%%%%%%%%%%%%%%%%%%%%%%%%%%%%%%%%%%%%%%%%%%%%%%%%
\begin{table}
 \centering
   \caption{Parameters of the four spectra in Figure~\ref{flux-var-fig}. obs1:
     pn spectrum on 2001; obs2: pn spectrum on 2011; obs2-l: pn spectrum from
     time-intervals with count rates $\le$0.75$\bar{x}$; obs2-h: pn spectrum
     from time-intervals with count rates $\ge$1.25$\bar{x}$. $\Gamma_{0.5-2}$:
     the 0.5-2 keV power law photon index. $R_{0.5keV}$: the data/model ratio at
     0.5 keV, the model is the extrapolation of the best-fit power law to the
     2-10 keV spectrum of obs2. R$_{f-10keV}$: the 2-10 keV flux ratio between
     other spectra and obs2, the absolute 2-10 keV flux for obs2 is
     $3.07{\times}10^{-12}~ergs~cm^{-2}~s^{-1}$; HR: hardness-ratio defined as
     (H-S)/(H+S), where S is the photon number in 0.5-2 keV and H is the photon
     number in 2-10 keV.}
     \begin{tabular}{@{}lccccc@{}}
\hline
   Spec  &  $\Gamma_{2-10}$ & $\Gamma_{0.5-2}$ & R$_{0.5keV}$ & R$_{f2-10}$& HR\\
\hline
   obs1   & 2.45$^{+0.12}_{-0.12}$ & 2.95$^{+0.64}_{-0.57}$ & 2.20$\pm$0.12 & 0.79 & -0.840\\
   obs2   & 2.37$^{+0.03}_{-0.03}$ & 3.16$^{+0.14}_{-0.14}$ & 2.64$\pm$0.03 & 1    & -0.853\\
   obs2-l & 2.21$^{+0.09}_{-0.08}$ & 3.15$^{+0.56}_{-0.50}$ & 2.40$\pm$0.11 & 0.72 & -0.847\\
   obs2-h & 2.39$^{+0.06}_{-0.06}$ & 3.18$^{+0.33}_{-0.31}$ & 2.92$\pm$0.08 & 1.29 & -0.863\\
\hline
    \end{tabular}
  \label{flux-var-tab}
\end{table}
%%%%%%%%%%%%%%%%%%%%%%%%%%%%%%%%%%%%%%%%%%%%%%%%%%%%%%%%%%%%%%%%%%%%%%%%%%%%%%%%%%%%%

\begin{figure*}
\begin{center}
\includegraphics[bb=54 216 558 600, clip=1,scale=0.8]{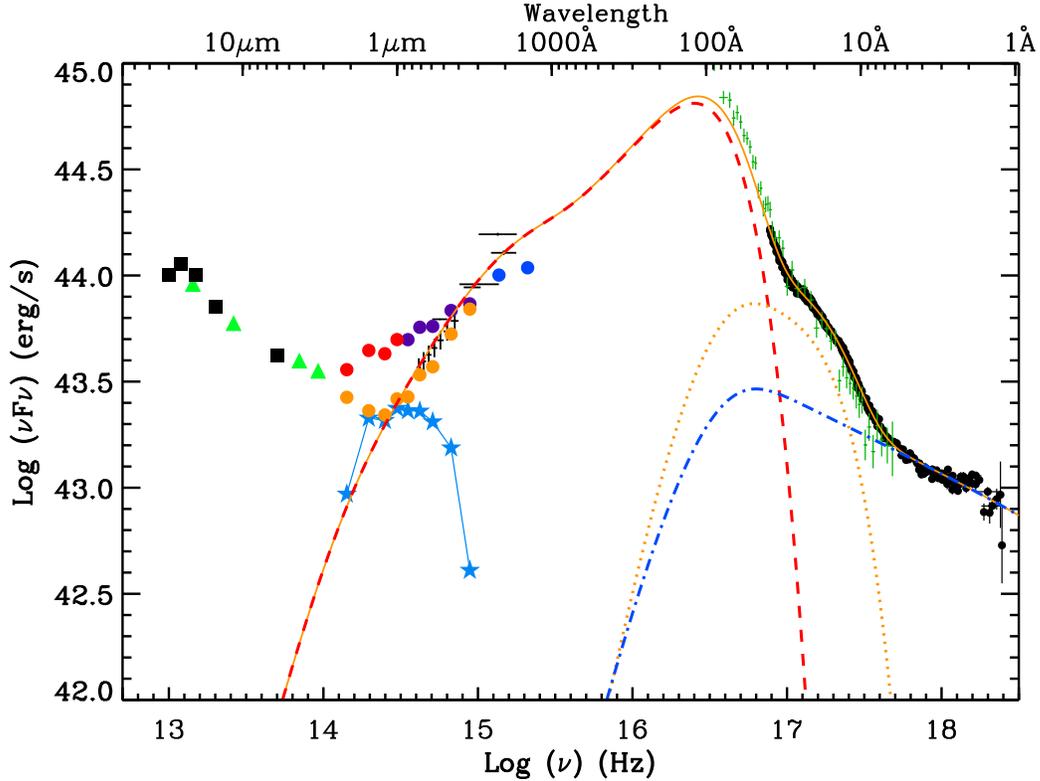}
\end{center}
\caption{The broadband SED of PG 2144+026, corrected for both Galactic and
  intrinsic reddening/extinction. Data includes (from high to low energy)
{\it XMM-Newton} EPIC spectra (black), 
{\it ROSAT} PSPCB spectrum (green) spectrum above
  100{\AA}); {\it GALEX} photometry (blue circles); {\it XMM-Newton}
  OM photometry (black), SDSS fiber spectrum (black)
  SDSS {\tt ugriz} photometry: PSF magnitude (purple
  circles), Petrosian magnitude (orange circles) and their discrepancy (cyan
  stars); {\it UKIDSS} YJHK photometry: Aper-3 magnitude (red circles),
  Petrosian magnitude (red circles) and their discrepancy (cyan stars); {\it
    WISE} 4-band photometry (green triangles); {\it Spitzer} IRS continuum
  (black squares). Orange solid line is the best-fit model to the {\it
    XMM-Newton} EPIC spectra, OM points, {\it ROSAT} spectrum and {\it SDSS}
  spectrum, using {\tt optxagnf} model in {\sc xspec} which includes accretion
  disc (red dashed line), low temperature Comptonisation (orange dotted line)
  and high temperature Comptonisation (blue dash-dotted line). Note that the
  cyan stars form a typical spectral shape of host galaxy contamination.}
\label{sed-fig}
\end{figure*}

\section{Multi-wavelength spectral energy distribution}

In this section, we construct the multiwavelength SED from
non-simultaneous observations.  Hence we first have to assess the effect of
long term variability.

\subsection{Long term variability} 

The {\it XMM-Newton} observation has provided high quality
simultaneous data in both X-ray and optical/UV for PG 2144+026. We
also made use of the {\it SDSS} spectrum to help define the optical
continuum. Although there was a 10 year interval between the {\it SDSS}
(obs-date: 2001-04-25) and {\it XMM-Newton} observation, we found that
they were fairly consistent with the OM optical data. There is a
previous XMM-Newton dataset taken in 2001-06-17 (60ks of good exposure
in every EPIC camera), close in time to the
SDSS data, but the OM was switched off for this observation. 

Instead we extract the X-ray data from the 2001 {\it XMM-Newton}
observation (blue: Fig~\ref{flux-var-fig}) 
and compare this with the time-averaged X-ray
spectra from our 2011 {\it XMM-Newton} data (black: Fig
\ref{flux-var-fig}).  While the flux at 10~keV remains almost the same,
the flux below 0.5~keV is lower by a factor 0.75 in the 2001
observation.  We compare this to the short timescale variability seen
within our single observation by accumulating a high flux spectrum
(green: where the flux is more than 25\% higher than the mean) and a
low flux spectrum (red: where the flux is more than 25\% lower than
the mean). Plainly the blue and red spectra are quite similar, so the
long timescale variability over decades is of similar amplitude to the
short timescale variability seen within a single observation. Hence it
seems likely that there is no dramatic change in the SED over the
timescales spanned by the multi-wavelength data. Details of these X-ray
spectra are given in Table \ref{flux-var-tab}.

There is also archival data from ROSAT PSPCB which can be used to extend the
X-ray spectrum down to 0.1-0.2~keV. This was taken on 1991-12-22. We
extracted this using {\tt XSELECT} v2.4b and followed the standard
procedure to reduce the data and extract the source and background
spectrum. This is also shown on the long term X-ray variability
figure, where it is a factor 0.44 below the time averaged 2011
XMM-Newton data although its shape is similar in the 0.3-2~keV region
of overlapping bandpass.

\begin{table*}
 \centering
  \begin{minipage}{175mm}

   \caption{The parameters of the {\sc optxagnf} model in Fig 12. nH$_{int}$ is the best-fit intrinsic extinction;
     $z$ is the redshift;
     R$_{cor}$ is the coronal radius; T$_{e}$ is the
     electron temperature for the soft X-ray Comptonisation; $\tau$ is the
     optical depth of the low temperature electron population; F$_{pl}$ is the
     fraction of luminosity contained in the hard X-ray Comptonisation compared
     to the total coronal luminosity. FWHM was measured from the H$\beta$ line
     profile after subtracting the narrow component.}
   \label{sed-spec-par-tab}
     \begin{tabular}{lccccccccccc}
\hline
Name  & nH$_{int}$ &	$z$ & $\Gamma_{2-10keV}$ & L$_{bol}$           &  M$_{BH}$        & L/L$_{Edd}$ & R$_{cor}$ & T$_{e}$ & $\tau$ & F$_{pl}$ & H$\beta$~FWHM \\
      & (10$^{+20}$) &           &                    & (erg s$^{-1}$)  & (M$_{\odot}$)
&             & (R$_{g}$) & (keV)   &        &     & (km s$^{-1}$) \\
\hline
PG 1244+026	& 3.12 & 0.048	&2.37$\pm$0.03	& 1.66$\times10^{45}$	&1.62$\times10^{7}$	&0.79	&12	&0.212	&16.9	&0.290	&940\\
\hline
   \end{tabular}
 \end{minipage}
\end{table*}

\subsection{Modelling the SED}

We use the broadband SED model for AGN ({\tt optxagnf} in {\sc xspec},
D12) to fit the optical/UV and X-ray data,
so as to recover the spectrum in the unobservable UV region due to the
inevitable Galactic extinction.  We allow a constant normalisation
offset between the {\it ROSAT} and 2011 {\it XMM-Newton} spectra due
to long-term variability, but assumed an identical spectral shape. The
broadband SED after corrected for both Galactic and intrinsic
reddening/extinction is plotted in Figure~\ref{sed-fig}, and best-fit
parameters listed in Table~\ref{sed-spec-par-tab}.

We also collected other archived data in optical, UV and infrared
wavelengths to build a more complete broadband SED. First we put the
{\it SDSS} {\tt ugriz} photometric points (PSF magnitude) on the SED
(purple circles in Figure~\ref{sed-fig}). These points appear higher than
the spectral data, which is mainly due to the inclusion of host galaxy
emission in the big aperture. Once we used the Petrosian
magnitudes, the aperture effect disappeared. The difference between PSF
magnitude and Petrosian magnitude was consistent with a typical host
galaxy spectrum (cyan stars in Figure~\ref{sed-fig}). Then we put YJHK
photometry points from {\it UKIDSS} LAS on the plot, including both
Aperture-3 magnitude (2\arcsec diameter, red circles in
Figure~\ref{sed-fig}) and Petrosian magnitude (obs-date: 2009-06-05).
The difference between these
two magnitudes is again due to the host galaxy contamination, which is
also consistent with {\it SDSS}. Other data points in
Figure~\ref{sed-fig} were from {\it GALEX} photometry (the two blue
circles in Figure~\ref{sed-fig}, obs-date: 2004-04-15), {\it WISE} 4
bands photometry (the four green triangles in Figure~\ref{sed-fig},
obs-date: averaged over 13 observations between 2010-01-17 and
2010-06-27) and {\it Spitzer} IRS continuum (the five black squares in
Figure~\ref{sed-fig}, obs-date: 2006-01-31; Veilleux et al., 2009).

The broadband SED of PG 1244+026 shows that the 
disc emission extends into the soft X-ray range, making the separate
soft X-ray component seen at the lowest X-ray energies. 
This connects smoothly to the (starlight subtracted) 
optical/UV disc emission, and is the
most prominent spectral component. 
The 1$\mu$m minimum feature also emerges after subtracting the
host galaxy (Landt et al. 2011). The spectrum rises towards the
infrared beyond 1$\mu$m, which is the standard signature of 
the inner region of dusty torus (the modeling of which is
beyond the scope of this paper). Although the data came from
various facilities at different observation times, they combine to 
give a smooth continuum. This implies that the optical and infrared
emission from PG 1244+026 is not strongly variable. 

The {\sc optxagnf} code of D12 connects the soft X-ray excess and hot
corona energetically to the cool disc. This takes as free parameters
the mass and spin of the black hole as well as the mass accretion
rate, parameterised as $L/L_{Edd}$. Besides these physical free
parameters, the model assumes that the flow thermalises to a colour
temperature corrected blackbody only down to a radius
$R_{cor}$ ($>R_{isco}$) so that the remaining accretion energy can power
the hot corona and cool soft excess components.  We set the spin to be
zero (see also the companion paper to this: Done et al. 2013, hereafter
D13) get a
best fit mass of $1.6{\times}10^{7}M_{\sun}$ for $L/L_{Edd}\sim 0.8$
and $R_{cor}=12R_g$. For this mass and mass accretion rate, the
standard disc emission already extends into the softest observable
X-ray band. Substantial spin overpredicts the observed soft X-ray
emission, so showing the potential to constrain black hole spin via
disc continuum fits in this object (D13).

We note that the best fit {\it optxagnf} model, together with blurred
{\it pexmon} reflection,  gives $\chi^2=2277/1801$ when fit to the
X-ray spectra used in Section 3, showing that this is a comparably
good fit to the data as the more phenomenological models where the
blackbody (inner disc) temperature is a free parameter.

\subsection{Independent estimates of Black Hole Mass}

\begin{figure}
\centering
\includegraphics[bb=0 0 576 625,width=8cm,height=9cm,clip=]{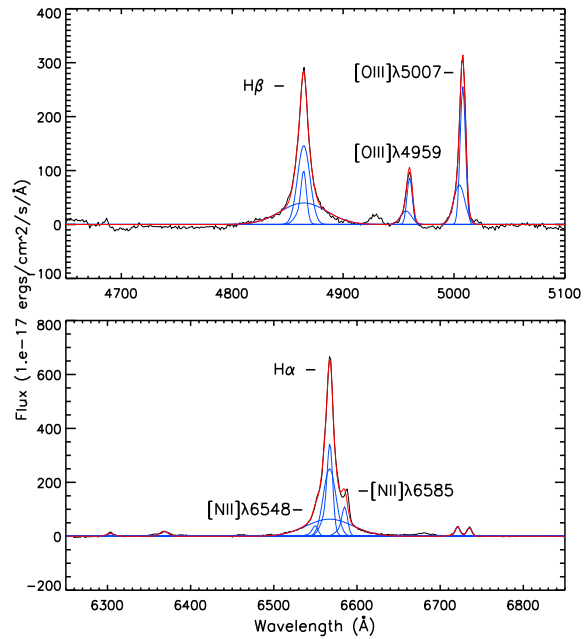}
\caption{The Multi-Gaussian fit to the nearby region of H$\alpha$ and H$\beta$
  line in the {\it SDSS} spectrum of PG 1244+026. The blue
  line shows Gaussian components separately, except for the narrow components
  and [NII]$\lambda$6585/6548 doublets which instead uses the same profile as
  the whole [OIII]$\lambda$5007 line. FeII lines have been subtracted (see Jin et al. 2012a)}
\label{Balmer-line-fig}
\end{figure}

PG 1244+026 has the narrowest Balmer line width among all PG quasars
with z $\le$0.5.  A three-Gaussian fitting to the H$\alpha$ and
H$\beta$ line found FWHM $=~810$~km~s$^{-1}$ and $3040$~km~s$^{-1}$ for
the intermediate and broad components, respectively. The direct FWHM
measurement on the combined profile was $940$~km~s$^{-1}$
(Figure~\ref{Balmer-line-fig}; see J12a and Jin, Ward \& Done (2012b)
for detailed line fitting
procedure).  The monochromatic luminosity at 5100{\AA} was
$4.52{\times}10^{43}$~ergs~s$^{-1}$. Therefore, using Equation 5 in
Vestergaard \& Peterson (2006), we obtained a black hole mass of
$4.8_{-1.2}^{+4.6} \times 10^6 M_{\sun}$. However, the implied high
$L/L_{Edd}$ means that radiation pressure can be important. Using
Equation 9 in Marconi et al. (2008) then gives a mass of
$1.8{\times}10^{7}M_{\sun}$, consistent with the mass from SED fitting.

Another way to estimate mass is via the excess variance. Ponti et
al. (2012) presented the correlation between hard X-ray excess
variance and black hole mass. These are tabulated for different lightcurve
lengths in Table \ref{excess-var-tab}. The 
40ks 2-10~keV excess variance (100s binned) is
$0.0338 \pm 0.0016$,  corresponding to $0.5 - 1.5 \times 10^7 M_\odot$ by
comparison to the reverberation mapped sample of Ponti et al. (2012). 
Therefore, our mass estimates from radiation pressure corrected
H$\beta$ FWHM, X-ray variability and SED fitting are all broadly
consistent.

% \begin{table*}
% \centering
%   \begin{minipage}{180mm}
%    \caption{Excess Variances (in percentage) of the lightcurves in Figure~\ref{lc-fig} for different segment lengths. To compare with Ponti et al. (2012), we also list the results for 250s binned lightcurves in 0.3-0.7keV, 0.7-2keV and 2-10keV bands.}
%    \label{excess-var-tab}
%      \begin{tabular}{ccccccc}
% \hline
% 	& $\sigma^{2}_{NXS,100s}$ & $\sigma^{2}_{NXS,100s}$ & $\sigma^{2}_{NXS,100s}$ & $\sigma^{2}_{NXS,250s}$ & $\sigma^{2}_{NXS,250s}$ & $\sigma^{2}_{NXS,250s}$\\
% Segment & 0.3-1 keV & 1-2 keV & 2-10 keV & 0.3-0.7 keV & 0.7-2 keV & 2-10 keV \\
% \hline
% 10ks	& $1.70\pm 0.03$ & $1.8\pm 0.06$ &
% 	$2.0 \pm 0.1$ & $1.7\pm 0.03$ &
% 	$1.7\pm 0.04$ & $1.8 \pm 0.1$ \\
% 20ks	& $2.90 \pm 0.04$ & $3.0\pm 0.07$
% 	& $2.8 \pm 0.1$ & $2.9\pm 0.04$  &
% 	$2.9\pm 0.05$ & $2.6\pm 0.1$ \\
% 40ks	& $3.60\pm 0.04$ & $3.6\pm 0.08$ &
% 	$3.4\pm 0.2$ & $3.5\pm 0.05$ &
% 	$3.5\pm 0.05$ & $3.1\pm 0.1$ \\
% 80ks	& $4.40 \pm 0.05$ & $4.3 \pm 0.1$
% 	& $3.9\pm 0.2$ & $4.4\pm 0.06$ &
% 	$4.3\pm 0.07$ & $3.6\pm 0.2$ \\
% \hline
%    \end{tabular}
%    \centering
%  \end{minipage}
% \end{table*}

\begin{table}
\centering
\caption{Excess Variances (in percentage) of the lightcurves in Figure~\ref{lc-fig} for different segment lengths. To compare with Ponti et al. (2012), we also list the results for 250s binned lightcurves in 0.3-0.7keV, 0.7-2keV and 2-10keV bands.}
\begin{tabular}{@{}ccccc@{}}
\hline
Seg.	   & 10ks & 20ks & 40ks & 80ks \\
\hline
$\sigma^{2}_{100s}$&&&&\\
0.3-1keV         &$1.74\pm 0.03$&$2.88\pm 0.04$&$3.56\pm 0.04$&$4.41\pm 0.05$\\
\hline
$\sigma^{2}_{100s}$&&&&\\
1-2keV           &$1.83\pm 0.06$&$2.98\pm 0.07$&$3.60\pm 0.08$&$4.30\pm 0.11$\\
\hline
$\sigma^{2}_{100s}$&&&&\\
2-10keV          &$2.05\pm 0.13$&$2.85\pm 0.15$&$3.38\pm 0.16$&$3.90\pm 0.20$\\
\hline
$\sigma^{2}_{250s}$&&&&\\
0.3-1keV         &$1.71\pm 0.03$&$2.84\pm 0.04$&$3.52\pm 0.04$&$4.36\pm 0.05$\\
\hline
$\sigma^{2}_{250s}$&&&&\\
1-2keV           &$1.73\pm 0.06$&$2.87\pm 0.07$&$3.49\pm 0.08$&$4.20\pm 0.10$\\
\hline
$\sigma^{2}_{250s}$&&&&\\
2-10keV          &$1.82\pm 0.11$&$2.61\pm 0.13$&$3.13\pm 0.14$&$3.62\pm 0.18$\\
\hline
\end{tabular}
\label{excess-var-tab}
\end{table}

\section{Summary}

In this paper, we presented a detailed spectral and timing analysis on
a 123ks {\it XMM-Newton} observation of PG 1244+026, a NLS1 with very
narrow Balmer lines and extreme mass accretion rate. The X-ray
spectrum of PG 1244+026 contains a very steep hard X-ray power law and
a strong/smooth soft X-ray excess. We show that the soft X-ray excess
can be fit with an additional cool Comptonisation component or
by extremely smeared, partially ionised reflection, but that both
models require additional blackbody emission at the softest X-ray
energies. 

We then use a detailed timing analysis to distinguish between these
two physical models for the soft X-ray excess.  We derive the
frequency-dependent fractional {\it RMS} and covariance spectra. For
the fastest variability, the {\it RMS} spectrum has a clear soft
excess, though this is smaller relative to the power law than in the
time averaged spectrum. However, the spectrum of the variability
correlated with the 4-10~keV lightcurve shows no soft excess over the
same timescales. This clearly shows that the majority of the soft
X-ray excess varies incoherently with the hard X-ray flux on fast
timescales ($<5000$~s). A similar drop in coherence between hard and
soft variability on the fastest timescales has previously also been
seen in NGC4051 (M$^{c}$Hardy et al. 2004).  Here, combining the variability
information with spectral models rules out a single reflection
component making both the soft X-ray excess and the iron K line
features as this predicts correlated flux variability in both hard and
soft bands. However, it does not rule out more complex models, where
there are multiple reflectors or some (small) contribution of ionised
reflection to the soft bandpass as well as an additional soft X-ray
component which emerges only below 1~keV.

By contrast, the spectrum of fast variability correlated with the
0.3-1~keV lightcurve has more soft X-ray excess than that of the HF
{\it RMS}, as it includes all the soft excess variability seen in the
HF {\it RMS} which is uncorrelated with the 4-10~keV emission.
However, the drop in HF 0.3-1~keV covariance below 0.5~keV confirms
the reality of the separate blackbody component required from the
spectral fits.

Thus the fast variability data are clearly consistent with the
spectral model in which the soft X-ray excess is a separate
Comptonised component which provides the seed photons for the higher
energy power law emission, but where the very softest energies are
dominated by the disc itself. Conversely, it is inconsistent with a
single reflection component producing the majority of the soft X-ray
excess and the 4-10~keV emission.

The slower variability does not show a clear difference between the
{\it RMS} and covariance spectra, showing that the source spectrum varies
coherently on timescales longer than 10~ks. This could be interpreted
in the spectral decomposition above in terms of propagation of
fluctuations from the disc, through the soft X-ray excess region and
into the high energy region (see e.g. Ar\'{e}velo \& Uttley 2006; Fabian
et al. 2009).

We also assemble a broadband SED for PG 1244+026 from far infrared to
hard X-ray, showing that this is dominated by the disc emission and
that this extends into the soft X-ray bandpass, consistent with the
spectral decomposition above. This system, and other NLS1 with similar
spectral properties (e.g. RE J1034+396, RX J0136.9-3510), can then be
used to constrain to the black hole spin via disc continuum fitting,
as high black hole spin over-predicts the observed soft X-ray flux
(D13).

\section*{Acknowledgements}
This work is mainly based on observations obtained with {\it XMM-Newton}, an ESA
science mission with instruments and contributions directly funded by ESA Member
States and NASA. This work makes use of data from {\it SDSS}, whose funding is
provided by the Alfred P. Sloan Foundation, the Participating Institutions, the
National Science Foundation, the U.S. Department of Energy, the National
Aeronautics and Space Administration, the Japanese Monbukagakusho, the Max
Planck Society, and the Higher Education Funding Council for England. We have
made use of the {\it ROSAT} Data Archive of the Max-Planck-Institut f\"{u}r
extraterrestrische Physik (MPE) at Garching, Germany.

\end{document}